\documentclass[12pt]{article}\usepackage[hyperfootnotes=false]{hyperref}
   \usepackage{epsfig}
      \usepackage{amsmath}
      \usepackage{amssymb}
  \usepackage{graphicx}
  \usepackage{color}
  \newlength{\abstractwidth}
\numberwithin{equation}{section}

\def\eps{\epsilon}
\def\p{\partial}
\def\vs{\vskip .1 in}
\newcommand{\bea}{\begin{eqnarray}}
\newcommand{\eea}{\end{eqnarray}}
\newcommand{\be}{\begin{equation}}
\newcommand{\ee}{\end{equation}}
\def\subsec{\subsection}
\def\M{{\cal M}}
\def\N{{\cal N}}
\def\A{{\cal A}}

\def\C{{\cal C}}

\def\cH{{\cal H}}
\def\L{\Lambda}
\def\Vol{{\rm Vol}}
\def\IH{\mathbb{H}}
\def\eqr{\eqref}
\def\b{\beta}
\def\l{\lambda}
\def\lang{\langle}
\def\rang{\rangle}
\def\N{{\cal N}}
\def\la{\langle}
\def\ra{\rangle}
\def\L{\Lambda}
\def\cH{{\cal H}}

\def\L{\Lambda}

\def\p{\partial}

\def\Oc{{\cal O}}
\def\half{{1\over 2}}
\def\rar{\rightarrow}

\def\a{\alpha}
\def\b{\beta}

\def\l{\lambda}
\def\t{\tau}
\def\eps{\epsilon}

\def\vs{\vskip .1 in}
\def\bul{$\bullet$~}

\def\IZ{\Bbb{Z}}

\def\Tr{{\rm Tr}}
\def\d{\delta}
\def\t{\tau}
\def\IR{\mathbb{R}}
\def\S{\Sigma}
\def\bul{$\bullet$ \quad}
\def\E{{\cal E}}
\def\X{{\cal X}}
\newcommand{\es}[2] {\begin{equation} \label{#1} \begin{split} #2 \end{split} \end{equation}}
\def\D{\Delta}
\def\L{{\cal L}}

\def\la{\label}

\begin{document}
\renewcommand{\theequation}{\thesection.\arabic{equation}}

\begin{titlepage}
  \rightline{}
  \bigskip

  \bigskip\bigskip\bigskip\bigskip

  \bigskip

 \centerline{\Large \bf { ~~ Universality in the geometric dependence of R\'enyi entropy }}
 
    \bigskip

  \begin{center}

 \bf { Aitor Lewkowycz$^1$ and Eric Perlmutter$^2$}
  \bigskip \rm
\bigskip

$^1$ Jadwin Hall,  Princeton University, Princeton, NJ 08544, USA
\\ ~~
\\
$^2$ DAMTP, Centre for Mathematical Sciences, \\University of Cambridge, CB3 0WA, UK  \\
\rm

\bigskip
\bigskip

\vspace{2cm}
  \end{center}

  \bigskip\bigskip

 \bigskip\bigskip
  \begin{abstract}

	We derive several new results for R\'enyi entropy, $S_n$, across generic entangling surfaces. We establish a perturbative expansion of the R\'enyi entropy, valid in generic quantum field theories, in deformations of a given density matrix. When applied to even-dimensional conformal field theories, these results lead to new constraints on the $n$-dependence, independent of any perturbative expansion. In 4d CFTs, we show that the $n$-dependence of the universal part of the ground state R\'enyi entropy for entangling surfaces with vanishing extrinsic curvature contribution is in fact fully determined by the R\'enyi entropy across a sphere in flat space. Using holography, we thus provide the first computations of R\'enyi entropy across non-spherical entangling surfaces in strongly coupled 4d CFTs. Furthermore, we address the possibility that in a wide class of 4d CFTs, the flat space spherical R\'enyi entropy also fixes the $n$-dependence of the extrinsic curvature contribution, and hence that of arbitrary entangling surfaces. Our results have intriguing implications for the structure of generic modular Hamiltonians.

 \medskip
  \noindent
  \end{abstract}

  \end{titlepage}

    \baselineskip=17.63pt \setcounter{footnote}{0}
  \tableofcontents

\section{Introduction}

It has become increasingly apparent that quantum entanglement is a profound aspect of quantum field theories. Given a reduced density matrix $\rho$, a natural observable is the R\'enyi entropy,
\begin{equation}\la{Renyi}
S_n=\frac{1}{1-n} \log \Tr \rho^n~,
\end{equation}
where $n$ is a non-negative integer. Upon analytic continuation of $n$ to the reals, the $n\rar 1$ limit yields the entanglement entropy, $S_{EE}=-\Tr\rho\log\rho$. We will be concerned with the degree of entanglement between two spatial regions $\A,\bar{\A}$ of a quantum system at fixed time, in which case $\rho$ is obtained by tracing over the degrees of freedom of $\bar{\A}$. 

We now know that entanglement entropy probes the fundamental structure of quantum field theories. For instance, suitably defined, it can be viewed as a dimension-independent measure of degrees of freedom in conformal field theories, which furthermore has a beautiful and robust geometric construction in the context of the AdS/CFT correspondence \cite{Ryu:2006ef}. The set of R\'enyi entropies, on the other hand, contains total information about the spectrum of $\rho$, yet its relations to fundamental aspects of quantum field theory and holographic spacetime are relatively poorly understood. The broad purpose of this paper is to strengthen both of these relations. 

In quantum field theory, one can compute R\'enyi entropy by passing to Euclidean time and inserting a conical deficit around the entangling surface $\S=\partial \A$ \cite{Callan199455}. The R\'enyi entropy will be given in terms of the partition function $Z_n$ of the geometry with a conical excess of $2\pi(n-1)$ across $\S$:
\begin{equation}\la{Renyi2}
S_n=\frac{\log Z_n-n \log Z_1}{1-n} ~.
\end{equation}
However, this conical singularity introduces a UV divergence, which we regulate by putting a cutoff $\epsilon$ away from the singularity. This results in the following general structure:
\begin{equation}\la{Renyidiv}
S_n=\frac{a_{d-2}(n)}{\epsilon^{d-2}}+\frac{a_{d-4}(n)}{\epsilon^{d-4}}+...+ a_{{\text{even}}}(n) \log \epsilon+a_0(n)+\ldots
\end{equation}
where the $\log$ term only appears in even dimensions. In even dimensions, in analogy with the entanglement entropy, the ``universal'' term of the R\'enyi entropy is the $\log \epsilon$ term, $a_{\rm even}(n)$; in odd dimensions, the universal term is the constant term $a_0(n)$. 

In practice, it is challenging to actually compute these functions. In this paper we are going to develop a ``R\'enyi perturbation theory'' in deformations of the reduced density matrix, applicable in general quantum field theories. This will be shown to boil down to computations of correlation functions of the deforming field theory operators on the aforementioned conical spacetime.

To see what this is good for, we will derive new {\it non-perturbative} results regarding the geometric dependence of R\'enyi entropy. The functions $a_i(n)$ in \eqr{Renyidiv} are implicit functions of the shape of the entangling surface, $\Sigma$. The shape-dependence of R\'enyi entropy is generally complicated, and often neglected in favor of highly symmetric (e.g. planar or spherical) entangling surfaces. An obvious and interesting application of our perturbative formalism is to the case of ``geometric perturbations'', where either the shape of $\S$ or the background geometry on which the quantum field theory lives is deformed. In this case, the deformation is controlled by the stress tensor. This type of perturbation was considered for the entanglement entropy in \cite{RS1}. The linearized perturbation of the R\'enyi entropy, for example, is given by the simple form
\be\la{fogeoint}
\d^{(1)}S_n = {n\over 2(1-n)}\int_{\M} \big(\lang T^{\mu\nu} \rang_{n} - \lang T^{\mu\nu} \rang_{1}\big) h_{\mu\nu}
\ee
where $h_{\mu\nu}$ is the geometric perturbation, $\M$ is the manifold on which the quantum field theory lives, and $\lang T^{\mu\nu} \rang_n$ is the stress tensor expectation value on the conical spacetime in the given state. Taking the $n\rar 1$ limit gives the entanglement results of \cite{RS1}.

If one focuses on geometric perturbations of the vacuum R\'enyi entropy in conformal field theories across a plane or a sphere, which are conformally equivalent configurations \cite{ch2}, $\lang T^{\mu\nu} \rang_{n}$ takes a simple form, and \eqr{fogeoint} can be explicitly evaluated for a given $h_{\mu\nu}$. The $n$-dependence of the perturbation can be written in the following tidy form:
\be\label{15}
\d^{(1)}S_n(\S=S^{d-2}) \propto {n\over 1-n}(S_{n=1}- S_n - (n-1) S'_n)\big|_{\S=S^{d-2}}
\ee
where $S'_n \equiv \p_nS_n$. (Recall that $S_{n=1}=S_{EE}$.)

Actually, \eqr{fogeoint} and \eqr{15} are even more powerful: for even-dimensional conformal field theories, they imply constraints on the R\'enyi entropy that are non-perturbative in the shape of $\S$ and the background geometry. To explain this, we first note that the universal pieces $a_{\rm even}(n)$ and $a_0(n)$ are further constrained in conformal field theories. This is especially true in even dimensions, where they are effectively determined by scale invariance to be local functionals of the geometry of $\Sigma$. In four dimensions, for example, the structure of this term is \cite{Fursaev:2012mp}
\begin{equation}\label{fursaevint}
S_n=\left (- \frac{f_a(n)}{2\pi} \int_{\Sigma} R_{\Sigma}-\frac{f_b(n)}{2\pi} \int_{\Sigma} \big( {K^a_{i j}} {K^a_{j i}} -\frac{1}{2} (K^a_{ i i})^2\big)+\frac{f_c(n)}{2\pi} \int_{\Sigma} {C^{ab}}_{ab} \right ) \log R/\eps ~.
\end{equation}
$R_{\S}$ is the Ricci scalar of $\S$, $K^a_{i j}$ is the extrinsic curvature in the transverse direction $x^a$, and ${C^{a b}}_{ab}$ is the Weyl tensor projected in the directions transverse to $\S$. The functions $f_i(n)$ are theory-dependent: $f_a(n)$ is related to the thermal entropy of the CFT on the hyperboloid, $S^1_n\times \IH^{d-1}$ with $\b=2\pi n$ \cite{Hung:2011nu}, but relatively little is known about $f_b(n)$ and $f_c(n)$. When $n=1$, $f_a(1)=a$ and $f_b(1)=f_c(1)=c$, and this formula reduces to Solodukhin's formula \cite{Solodukhin:2008dh} for entanglement entropy.\footnote{In our conventions, a real scalar has $a=\frac{1}{360},c=\frac{1}{120}$.}

We will show that \eqr{fogeoint} and \eqr{15} imply that $f_a(n)$ and $f_c(n)$ are not independent functions: rather,
\begin{equation} \label{eq:fca}
f_c(n)=\frac{n}{n-1} (a-f_a(n)-(n-1) f'_a(n))~.
\end{equation}
Equivalently, $f_c(n)$ can finally be given an {\it a priori} definition: namely, it is proportional to the expectation value of the stress tensor in the conical background. The latter was studied for free theories in \cite{Frolov:1987dz}. Analogous results apply in all even-dimensional CFTs.

Among other applications to be discussed, the relation \eqr{eq:fca} and its higher-dimensional analog allows us to perform the first holographic computations of R\'enyi entropy in strongly coupled even-dimensional CFTs across non-spherical entangling surfaces: the function $f_a(n)$ can be computed via hyperbolic black hole entropy in a given bulk theory \cite{Hung:2011nu}, and passed through \eqr{eq:fca} to determine $f_c(n)$. At $n=1/2$, this result can also be framed as a computation of holographic bipartite logarithmic negativity \cite{Rangamani:2014ywa}. 

Similarly, using the second-order geometric perturbations, one can also define $f_b(n)$ in terms of integrated stress tensor two-point functions on the conical spacetime. One would like to test the recent conjecture that $f_b(n)=f_c(n)$ for all CFTs. This conjecture was based on the behavior at $n=1$, and a lattice calculation for the free conformal scalar and fermion \cite{Lee:2014xwa} that produced this result. Carrying the second order calculation to its end, however, is rather complicated, due to features of conformal field theory on the cone. We are not able to prove that $f_b(n)=f_c(n)$, but we develop arguments supporting this conjecture for a large class of CFTs, which includes the free fields, $\N=4$ super-Yang-Mills (SYM), and CFTs with classical gravitational duals.

In particular, this implies that knowing hyperbolic black hole entropy in type IIB supergravity is sufficient to compute the strong coupling behavior of R\'enyi entropy across {\it any} entangling surface in $\N=4$ super-Yang-Mills and its counterparts. 

These results have further intriguing consequences. It is natural to ask whether there is a feasible holographic prescription for computing R\'enyi entropy for non-spherical surfaces directly, e.g. involving some deformations of hyperbolic black holes. We comment on the challenges to such an approach and propose ``deformed hyperboloid'' geometries that generalize the conformal mapping of Casini, Huerta and Myers to non-spherical entangling surfaces. We also discuss the question of locality of the modular Hamiltonian for generic surfaces. We argue that, at least for CFTs in flat space, the sphere is the only entangling surface for which the modular Hamiltonian is local. By expanding R\'enyi entropy around $n=1$ (in the spirit of \cite{Perlmutter:2013gua}), our results imply that even non-local modular Hamiltonians have correlators that are largely determined by correlators of the CFT stress tensor, as we explicitly show.

The remainder of the paper is organized as follows. In Section 2, we develop the R\'enyi perturbation theory. In Section 3, we analyze the consequences of first order perturbations for R\'enyi entropy in even-dimensional CFTs for generic entangling surfaces, including the result \eqr{eq:fca}. Section 4 discusses some applications of our results, including holographic calculations for non-spherical entangling surfaces. Section 5 studies the second order perturbations in the general context of CFT on the cone. In Section 6, we comment on the nature of the modular Hamiltonian and the use of conformal transformations in computing R\'enyi entropy for generic surfaces. We close in Section 7 with an open-ended discussion. Appendices contain some computational details and peripheral material.

\section{R\'enyi perturbation theory}
We begin with some basic definitions to establish notation. We start with a $d$-dimensional Euclidean quantum field theory (QFT) living on some manifold $\M$ with metric $g_{\mu\nu}$, and foliate the spacetime along the direction of a coordinate acting as Euclidean time. On a fixed time slice, we choose a spatial region, $\A$, bounded by a $(d-2)$-surface, $\Sigma=\p \A$. We then form the reduced density matrix, $\rho$, obtained by tracing over all degrees of freedom living in the complement of $\A$; the R\'enyi entropy, $S_n$, is defined as in \eqr{Renyi}.

We take $\rho$ to be normalized, $\Tr\rho=1$. The dependence of $\rho$ on the data ($\M, g_{\mu\nu}, \Sigma$) and on the state is left implicit.  Because $\rho$ is a positive Hermitian operator, one is free to define the modular Hamiltonian, $K$, as follows:
\be
\rho = {e^{-K}\over \Tr e^{-K}}~.
\ee
We will say more about this object later. 

The reduced density matrix $\rho$ also has a path integral representation. One can write
\be
\Tr\rho^n ={Z_n\over Z_1^n}
\ee
where $Z_n$ is the partition function of the QFT on the $n$-fold covering space of $\M$, denoted $\M_n$. In terms of $Z_n$, the R\'enyi entropy is given in \eqr{Renyi2}.
The ``replica manifold'' $\M_n$ is endowed with a $\IZ_n$ replica symmetry. In the language of density matrices, the replica symmetry is manifest as cyclicity of the trace. 

We now initiate a perturbative expansion of the R\'enyi entropy around a given reduced density matrix in each of these pictures. We then specialize to geometric perturbations, which probe the shape ($\Sigma$) and background geometry $(g_{\mu\nu}$) dependence of the R\'enyi entropy.

\subsec{R\'enyi perturbation theory I: Density matrix}
Consider an infinitesimal perturbation of the reduced density matrix,
\be
\rho = \rho_0(1+ \d\rho)~.
\ee
Normalization of $\rho$ implies $\Tr(\rho_0 \d\rho)=0$. We wish to expand $\Tr\rho^n$. Before doing so, it is convenient to write $\Tr\rho^n$ as
\be
\Tr\rho^n = \Tr\prod_{j=1}^n\rho_j
\ee
where $j=1,\ldots n$ indexes replicas. This becomes important in perturbation theory, where there are independent correlations among replicas that cannot be related by replica symmetry. Accordingly, we denote the perturbation of the density matrix living on the $j$'th replica as $\d\rho_j$. To second order, the change in $\Tr\rho^n$ is
\be\la{ra}
\delta (\Tr\rho^n) \approx  n\Tr(\rho_0^n\d\rho)  +  \sum_{j< k}\Tr(\d\rho_j \rho_0^{k-j}\d\rho_k\rho_0^{n-k+j})+O(\d\rho^3)~.
\ee

We have used replica symmetry to simplify these expressions. Its effect on the linear term is to reduce it to $n$ times a one-point function of $\d\rho$ on $\M_n$, where $\d\rho$ lives on any {\it fixed} replica, that is,
\be\la{1pt}
\sum_{j=1}^n \Tr(\rho_0^n \d\rho_j) = n \Tr(\rho_0^n \d\rho_1)~.
\ee
The quadratic perturbation, however, is more subtle because of the fact that $\rho$ is an operator. Each term in the quadratic perturbation can be viewed as a two-point function between perturbations living on different replicas. We will make this interpretation more transparent in the path integral picture; for the time being, let us simply write the linearized perturbation of the R\'enyi entropy, $\d^{(1)}S_n$, using \eqr{Renyi}:
\be
\d^{(1)} S_n = {n\over 1-n}{\Tr(\rho_0^n \d\rho_1)\over \Tr\rho_0^n}~.
\ee
That is, $\d^{(1)}S_n$ is proportional to a one-point function of $\d\rho$ in the replicated space $\M_n$, defined with respect to the reduced density matrix $\rho_0$. We write this as
\be\la{renyifirstpi}
\d^{(1)} S_n = {n\over 1-n}\lang \d\rho_1\rang_{n}~.%
\ee

\subsec{R\'enyi perturbation theory  II: Path integral on $\M_n$}

Consider a manifold $\M$ comprised of a $\tau$ circle times another manifold, $M$. The simplest example is the hyperboloid, $\M= \cH^d \equiv S^1\times \IH^{d-1}$ where the $S^1$ coordinate $\t$ has period $\t\sim \t+2\pi$, but this applies to more general geometries with a periodic coordinate.
For a general theory with fields $\phi$, we can write its action $I$ in terms of these $\tau$ coordinates: $I= \int_0^{2\pi} d\tau {\cal L}[\phi(\tau)]$. Now we can implement the replica trick by changing the radius of the circle: $\tau \sim \tau +2\pi n$ while keeping the metric periodic with $\tau \sim \tau+2\pi$. 
 
In other words, we have $n$ replicas with a $\mathbb{Z}_n$ replica symmetry among them. The partition function will just be given by the path integral in the multicovered space:
\begin{equation}\la{Zn}
Z_n=\int_{\phi(2\pi n)=\phi(0)} {\cal D} \phi ~e^{-I_n[\phi]}~.
\end{equation}  

When this $\tau$ circle is an isometry of the metric, this is just a finite temperature partition function with $\b=2\pi n$. This reconnects easily with the usual density matrix approach to computing R\'enyi entropy
\bea
-n \partial_n \log Z_n=-\int_{S^1_{n} \times M} \langle T_{\tau \tau} \rangle_n=-2 \pi n \int_{M} \langle T_{\tau \tau} \rangle_n=  n \langle K \rangle_n~.
\eea
That is, $ Z_n=\int {\cal D} \phi e^{-I_n}=\Tr e^{-n K}  $. If $M$ doesn't have a $\t$ isometry, one cannot directly identify the modular Hamiltonian in this manner because the integrand will depend on $\tau$. We have presented the argument with an explicit circle for simplicity, but, of course, one can also apply it to other geometries like a cone; there we also have a (warped) $\tau$ circle. 

Now we can do perturbation theory from this path integral approach rather easily: we want to compute 
\be
\d S_n = {1\over 1-n}(\d \log Z_n - n\d \log Z_1)~.
\ee
We perturb the action $I_n$, and expand. Let us denote this perturbation as

\be\la{actpert}
\delta I_n = -\int_{\M_n}  g O
\ee
where $g$ is a perturbative coupling and $O$ is a QFT operator. Using \eqr{Zn}, the perturbative expansion of $\log Z_n$ is
\begin{eqnarray}
\delta \log Z_n
=g n \int_{\M_1} \langle O \rangle_n +\frac{g^2}{2} n   \int_{\M_1} \int_{\M_n} \langle O O \rangle_{n,c}+O(g^3)~. \label{eq:Renyiexp}
\end{eqnarray}
The ``$c$'' subscript indicates a connected correlator, on account of the log. 

Note that we have used replica symmetry to say that\footnote{From the CFT point of view, the one-point function is periodic with $\tau \sim \tau+2\pi$.}
\be\int_{\M_n} \langle O \rangle_n=n \int_{\M_1} \langle O \rangle_n~.
\ee
This is equivalent to \eqr{1pt}. However, when we deal with higher order correlators appearing in the $O(g^2)$ term and beyond, we cannot pull out another factor of $n$ because there is interaction among replicas: two-point functions of $O$ depend nontrivially on the spacing between their respective copies. Analogous considerations apply, for instance, in the classification of operators of $\mathbb{Z}_n$ orbifold CFTs.

This expansion may seem deceptively simple, but let us compare the second order perturbation in \eqr{eq:Renyiexp} with the analogous term in the density matrix approach. (For simplicity, we temporaily consider the unnormalized reduced density matrix, which we call $\tilde\rho$.) Normally when one talks about second order perturbations of the R\'enyi entropies by thinking about a QFT in the replicated space $\M_n$, two types of correlations appear: one coming from expanding each density matrix to second order, and the other from expanding two different density matrices to first order. If we introduce the notation $\Oc_j$ as the integrated operator over the $j$'th replica,
\be
{\cal O}_j =\int_{2\pi (j-1)}^{2\pi j} d\tau O
\ee
then the unnormalized perturbation $\d\tilde\rho_j$ on replica $j$ is
\be\la{rhopert}
\delta \tilde\rho_j=g {\cal O}_j+\frac{g^2}{2} {\cal O}_j^2+O(g^3)~.
\ee
Expanding $\log \Tr (\tilde\rho+\delta \tilde\rho)^n$, one finds a second order contribution 
\es{denspert}{
\log \Tr (\tilde\rho+\delta \tilde\rho)^n|_{g^2}&=\frac{g^2}{2} \sum_j \langle {\cal O}_j {\cal O}_j \rangle_{n,c}+\frac{g^2}{2} \sum_{k \not = j} \langle {\cal O}_j {\cal O}_k \rangle_{n,c}   \\
&=\frac{g^2}{2} n \langle {\cal O}_1  {\cal O} \rangle_{n,c}~.
}
We have used the replica symmetry to fix $j=1$, and the notation
\be
\Oc =  \sum_{j = 1}^{n}  {\cal O}_j =\int_0^{2 \pi n} d\tau O=\int_{\M_n} O
\ee
to denote the operator integrated over the full space $\M_n$. The result \eqref{denspert} is equivalent to the second order result \eqr{eq:Renyiexp} from the path integral.

To summarize, the linear and quadratic perturbations of the R\'enyi entropy induced by an action perturbation \eqr{actpert} are 
\es{sum}{
\d^{(1)}S_n &= {n\over 1-n} \left(\int_{\M_1} \langle O \rangle_n -  \int_{\M_1}\langle O \rangle_1\right)  \\
\d^{(2)}S_n &=  {n\over 2(1-n)}\left( \int_{\M_1} \int_{\M_n} \langle O O \rangle_{n,c} -  \int_{\M_1} \int_{\M_1} \langle O O \rangle_{1,c} \right)~.
}
In Appendix B, we make some comments about how to extract the correct $n\rar1$ results from these expressions. 

\subsec{First order geometric perturbations}
Having established a general framework for R\'enyi perturbation theory, we begin study of the first order perturbation in \eqr{sum}. 
In particular, we specialize to {\it geometric} perturbations \cite{RS1}. These are defined as those $\d\rho$ which are induced by either a shape deformation of $\Sigma$, or a metric deformation of the background geometry $g_{\mu\nu}$.  One can write the metric near $\Sigma$ in ``adapted'' coordinates, such that this metric -- call it $\bar g_{\mu\nu}(\S)$ -- is a function of $g_{\mu\nu}$ and the shape of $\Sigma$. Then geometric perturbations are conveniently packaged as perturbations of the adapted metric,
\be
g'_{\mu\nu}(\S) =\bar g_{\mu\nu}(\S)+h_{\mu\nu}(\S)~.
\ee
Henceforth, we drop the explicit $\S$ dependence of the metric. 

As shown in \cite{RS1}, for such perturbations, 
\be\label{221}
\d\rho = \half \int_{\M} \big( T^{\mu\nu}  - \lang T^{\mu\nu}\rang_{1}\big)h_{\mu\nu}
\ee
where the stress tensor is defined as $T^{\mu\nu}=- {2\over \sqrt{g}} {\d {\cal L} \over\d g_{\mu\nu}}$. This is to be understood as sitting inside a path integral. Indeed, \eqr{221} is quite clear from the path integral approach above, whereupon one simply chooses the deforming operator
\be\la{Th}
O = \half T^{\mu\nu} h_{\mu\nu}~.
\ee
Plugging into the first line of \eqr{sum} yields
\be\la{fogeo}
\d^{(1)}S_n = {n\over 2(1-n)}\int_{\M} \big(\lang T^{\mu\nu} \rang_{n} - \lang T^{\mu\nu} \rang_{1}\big) h_{\mu\nu}~.
\ee
This is a simple and interesting result: $\d^{(1)}S_n$ is fixed by the difference in stress tensor expectation values between the replicated and original QFTs. Taking the limit $n\rar 1$ reproduces the entanglement result of \cite{RS1}. 

To actually compute \eqr{fogeo}, one needs an explicit expression for the adapted metric perturbation $h_{\mu\nu}$. Following the notation of \cite{RS1}, we use $\lbrace x^a\rbrace, a=1,2$ to denote the two coordinates transverse to $\S$, and $\lbrace y^i \rbrace, i=1\ldots d-2$ to denote the coordinates along $\S$. Then to $O(x^2)$ in the distance from $\S$, the adapted metric is
\begin{eqnarray}
 d s^2&=& (\delta_{ab}-{1\over 3} R_{a c b d}|_\Sigma x^{c}x^{d})dx^a dx^b 
 + \left(A_i+{1\over 3}x^b \varepsilon^{d e} R_{i b d e}\big|_{\Sigma} \right)\varepsilon_{a c}\, x^a dx^c dy^i
  \label{metricexp}\\
 &+& \Big (\gamma_{ij}+2 K_{a ij} \, x^a+  x^a x^{c}\big( \delta_{a c} A_i A_j+R_{i a c j}|_\Sigma +K_{c i l} K_{a\,j}^{~l}  \big) \Big)dy^idy^j + \mathcal{O}(x^3) \nonumber
\end{eqnarray}
$\gamma_{i j}$ is the induced metric of the surface; $K^a_{i j}$ the extrinsic curvature corresponding to direction $a$; $R_{a b c d}$ the Riemann tensor evaluated on $\Sigma$; $\varepsilon$ is the induced volume form in the transverse space; and $A_i$ is a KK-like vector that comes from the $g_{i a}$ components. We use the conventions of \cite{RS1} where further definitions are provided. This foliation of the metric near the entangling surface was particularly useful in \cite{Lewkowycz:2013nqa,Dong:2013qoa,Camps:2013zua} to justify the prescription of \cite{Ryu:2006ef,Hung:2011xb}.

\subsec{Symmetric entangling surfaces in flat space CFTs}

Our results so far apply to QFTs with general $(\M,\Sigma,g_{\mu\nu})$. Let us specialize to the case of CFTs in flat space, $\M=\IR^d$, with a planar entangling surface. In such a case, \eqr{fogeo} simplifies even further. The covering space is the conical spacetime $\M_n = \C_n \times \IR^{d-2}$, where $\C_n$ has a conical excess $2\pi(n-1)$. We can write its metric as
\be\la{conemetric}
ds^2_{\C_n\times \IR^{d-2}} = dr^2 + r^2d\tau^2 + \d_{ij}dy^idy^j~,\quad \tau\sim \tau+2\pi n~.
\ee
$\Sigma$ sits at $r=0$ along $\lbrace y^i\rbrace$, and $(x^1,x^2)=(r\cos\t,r\sin\t)$. Tracelessness and maximal symmetry fix the stress tensor to take the diagonal form
\bea\la{FnR}
\lang {T^{\t}}_{\t} \rang_{\C_n \times \IR^{d-2}} &=& {F(n)\over r^d}\nonumber \\
\lang {T^{\mu}}_{\nu} \rang_{\C_n \times \IR^{d-2}} &=& -{ {\delta^{\mu}}_{\nu}\over d-1}{F(n)\over r^d}~ \quad (\mu, \nu \neq \t)
\eea
where $F(n)$ is a theory-dependent function that vanishes linearly as $n\rar 1$. Thus, we see from \eqr{fogeo} that $\d^{(1)}S_n$ is proportional to the energy density ${T^{\t}}_{\t}$ on $\C_n \times \IR^{d-2}$:
\be\la{flatgeo}
\d^{(1)}S_n(\S={\rm plane}) = {nF(n)\over 2(1-n)}\int_{\M} {1\over r^d}~\big( {h^{\t}}_{\t}-{1\over d-1}\sum_{i=1}^{d-1} {h^i}_i\big)
\ee
where $\int_{\M}$ is now shorthand for
\be
\int_{\M} \equiv \int_{\S} d^{d-2}y\int_0^{2\pi} d\t \int_0^{\infty} dr ~r
\ee

As is well-known, a planar entangling surface embedded in flat space can be conformally mapped to various other configurations, including to a spherical entangling surface in flat space \cite{ch2}. The reduced density matrix can also be conformally mapped to a thermal density matrix of the CFT living on the hyperboloid, ${\cal H}^d_n\equiv S^1_n\times \IH^{d-1}$, at inverse temperature $\b=2\pi n$. (We set the length scale of ${\cal H}^d_n$ to one.) The mapping from the conical metric \eqr{conemetric} to ${\cal H}^d_n$ in Poincar\'e coordinates is particularly simple,
\be
ds^2_{{\cal H}^d_n} = {1\over r^2} ds^2_{\C_n \times \IR^{d-2}}  = d\t^2 + ds^2_{\IH^{d-1}}~, \quad \t\sim \t+2\pi n~.
\ee
Under this map, $\rho$ and $K$ are unitarily equivalent to $\rho_T$ and $H$, the thermal density matrix and Hamiltonian on ${\cal H}^d_n$, respectively,
\bea
\rho &=& U^{-1} \rho_T U \nonumber \\
K &=& U^{-1}2\pi H U
\eea
and the partition function $Z_n$ becomes thermal. The stress tensor is constant on ${\cal H}^d_n$,
\bea\la{F(n)}
\lang {T^{\t}}_{\t}\rang_{{\cal H}^d_n}-\lang {T^{\t}}_{\t}\rang_{\cH^d} &=& F(n)\nonumber \\
\lang {T^{\mu}}_{\nu} \rang_{{\cal H}^d_n}-\lang {T^{\mu}}_{\nu} \rang_{\cH^d} &=& -{\delta^{\mu}}_{\nu}{F(n)\over d-1}~, \quad  \mu,  \nu \neq \t
\eea
where $\cH^d\equiv \cH^d_{n=1}$. Note that $\lang  {T^{\mu}}_{\nu} \rang_{\cH^d}\neq 0$ in even dimensions. 

We can write equation \eqr{F(n)} in yet another way that is particularly useful. The energy density on ${\cal H}^d_n$ is given by the usual thermodynamic relation $\p_{n}\log Z_n=2\pi \Vol(\IH^{d-1})\lang {T^{\t}}_{\t}\rang_{{\cal H}^d_n}$.
We can now use this relation to take derivatives of $(1-n) S_n$ and express it in terms of the one-point function of the stress tensor. Using the path integral definition of the R\'enyi entropy across the sphere, this can be re-written as
\be\la{fs}
2\pi \Vol(\IH^{d-1}) F(n) =(S_{n=1}- S_n - (n-1) S'_n)\big|_{\S=S^{d-2}}
\ee
where $S'_n \equiv \p_nS_n$. (Recall that $S_{n=1}=S_{EE}$.) We have thus shown that the first order geometric perturbation of the R\'enyi entropy across a plane, defined in \eqr{flatgeo}, is a linear function of the unperturbed R\'enyi entropy across a sphere and its first $n$-derivative. As we exploit in the next section, this result has powerful implications in even-dimensional CFTs. 

We make two related comments before moving on: \vs

1. The function $F(n)$ was studied many years ago in the rather different context of cosmic strings \cite{Frolov:1987dz}. The relation \eqr{fs} forges a 
 relationship between spherical R\'enyi entropy and stress tensors in cosmic string backgrounds. \eqr{fs} follows trivially from the definition of $F(n)$ in \eqr{FnR} by a conformal transformation; \cite{Dowker:2010bu} used it to compute the entanglement entropy (EE), but we would like to point out that one can also extract from it the whole R\'enyi entropy. To our knowledge, $F(n)$ has only previously been computed in a handful of free CFTs in various dimensions. 

\vs
2.  Note that $F(n)$ is basically the same thing as what \cite{Hung:2011nu} calls the dimension of a spherical twist operator. That is, one can think of the conical background as being generated by the insertion of a twist operator along $\S$: for some QFT operator $O$,
\be\la{twist}
\langle O \rangle_n={\langle O e^{-(n-1) K} \rangle_1\over \langle e^{-(n-1) K} \rangle_1}
\ee
where $K$ is the modular Hamiltonian associated to $\rho$. When $O=T$, one can interpret the one-point function of the stress tensor in the conical background as the scaling dimension of the twist operator, $h_n$. For a spherical entangling surface, $h_n=\frac{2\pi n}{d-1} F(n).$\footnote{Note that there is a factor of $n$ because \cite{Hung:2011nu} considers the stress tensor to be summed over replicas while we are putting it at a point of the replicated space.} Using \cite{ch2}, \cite{Hung:2011nu} computed $h_n$ for spherical twist operators (and hence $F(n)$) at strong coupling using holography. See \cite{Hung:2014npa} for recent progress regarding twist operators.

\section{R\'enyi entropy for arbitrary shapes in even-dimensional CFTs}

In the vacuum of even-dimensional CFTs, the universal, logarithmic term in the R\'enyi entropy is constrained to take the following schematic form:
\be\la{Renyischem}
S_n|_{\log} = \sum_i f_i(n) g_i(\S,g_{\mu\nu}) \log R/\eps~.
\ee
The functions $g_i(\S,g_{\mu\nu})$ are local integrals over $\S$ of conformally invariant combinations of curvatures, and the theory-dependent functions $f_i(n)$ contain the full $n$-dependence. Neither the local form of $S_n$, nor the factorization of shape dependence and $n$-dependence, holds for the finite term of odd-dimensional CFTs. 

Perhaps surprisingly, the form \eqr{Renyischem} can be used in conjunction with our perturbative approach to powerful effect: by introducing geometric perturbations around the sphere in flat space, we can bootstrap our perturbative first order results to determine relations among the $f_i(n)$, which were previously thought to be independent functions. Such relations hold in full generality. 

We start in the vacuum of a 4d CFT, where we show one of our main results: given \eqr{fursaevint}, the function $f_c(n)$ is fixed in terms of $f_a(n)$. That is,
the $n$-dependence of the R\'enyi entropy across an arbitrary entangling surface $\S$ which does not turn on $f_b(n)$ is fully fixed by the R\'enyi entropy across $\S=S^2$. We then show similar results in 6d CFTs, and provide a new explanation of why the R\'enyi entropy takes the form \eqr{Renyischem} in general even dimensions. 

In what follows, it will prove handy to have the explicit expression for the regulated hyperbolic volume: 
\be
\Vol({\mathbb{H}^{d-1}}) =  \frac{\pi^{\frac{d}{2}}}{\Gamma (\frac{d}{2})}
\times \left\{ 
\begin{array}{cc}
(-1)^{\frac{d-1}{2}} \,, & \text{$d$ odd} \\
(-1)^{\frac{d}{2}-1} 2\pi^{-1} \log R/\epsilon \,, & \text{$d$ even} \,.
\end{array}
\right.
\ee

\subsection{$d=4$}
For convenience, we briefly review what is known about 4d CFT vacuum R\'enyi entropy. 

\subsubsection{Review}
Consider the formula \eqr{fursaevint} for the universal part of the 4d CFT R\'enyi entropy in vacuum, which we repeat here:
\begin{equation}\label{fursaev}
S_n=\left (- \frac{f_a(n)}{2\pi} \int_{\Sigma} R_{\Sigma}-\frac{f_b(n)}{2\pi} \int_{\Sigma} \big( {K^a_{i j}} {K^a_{j i}} -\frac{1}{2} (K^a_{ i i})^2\big)+\frac{f_c(n)}{2\pi} \int_{\Sigma} {C^{ab}}_{ab} \right ) \log R/\eps 
\end{equation}

When $\S=S^2$, only the first term contributes, leaving
\be\la{sfa}
S_n(\S=S^2) = -4f_a(n)\log R/\eps = {2\over \pi}f_a(n)\Vol(\IH^3)~.
\ee
In this case, the modular Hamiltonian, call it $K_{\rm Sph}$, is local: 

\be\la{ksph}
K_{\rm Sph}=-2\pi \int_{\IH^{3}}T_{\tau \tau}~.
\ee
As we saw earlier, it can be conformally mapped to the thermal Hamiltonian of the CFT living on $\cH^4_n$. 

Near $n=1, f_a(n)$ behaves as \cite{Perlmutter:2013gua}
\be\la{faexp}
f_a(n)\approx a-{c\over 2}(n-1)+\ldots
\ee
where $(a,c)$ are central charges defined via the conformal anomaly,
\be
\lang {T^{\mu}}_{\mu} \rang = -{a\over (4\pi)^2} E_4 + {c\over (4\pi)^2} C_{\mu\nu\rho\sigma}C^{\mu\nu\rho\sigma}
\ee
where $E_4$ is the Euler density. To derive \eqr{faexp}, one uses \eqr{ksph} in conjunction with the general result \cite{Perlmutter:2013gua} that the expansion of $S_n$ near $n=1$, for any entangling surface, is nothing but an expansion in connected correlators of $K$,\footnote{There are subtleties in this expansion related to the proper definition of $K$, even for the sphere. For an example of this in the context of the 4d free conformal scalar, see \cite{Lee:2014x}.}

\be\la{correxp}
S_n  = S_1 +  \sum_{m = 1}^\infty {(-1)^{m} \over (m+1)!} \langle  \underbrace{K\ldots K}_{m+1} \rangle_{1,c}~ (n-1)^m \,.
 \ee
So in the expansion of $f_a(n)$, terms of $O((n-1)^m)$ are fixed by connected $(m+1)$-point functions of $T_{\t\t}$ on $\IR^4$.

At arbitrary $n$, $f_a(n)$ has been computed for free scalars, fermions and vectors using heat kernel methods on $\cH^4_n$ \cite{Casini:2010kt,Dowker:2010bu,Fursaev:2012mp}.
It has also been computed for 4d CFTs at strong coupling dual to Einstein gravity and higher derivative corrections thereof \cite{Hung:2011nu}, by using the conformal mapping of \cite{ch2} and AdS/CFT computations of hyperbolic black hole entropy. Otherwise, little is known about $f_a(n)$. 

Even less is known about the functions $f_b(n), f_c(n)$. The former is isolated by considering a cylindrical entangling surface, $\S=S^1\times \IR$; then
\be\la{cyl}
S_n(\S=S^1\times \IR) = -\frac{\ell}{2 R} f_b(n)\log R/\eps~,
\ee
with $\ell,R$ the length and radius of the cylinder, respectively. At $n=1$, $f_b(1)=f_c(1) = c$. Higher order terms in an expansion around $n=1$ are determined in principle by \eqr{correxp}, but cannot be (and have not been) computed without detailed knowledge of $K$ for non-spherical $\S$. For general $n$, $f_c(n)$ has only been computed for free scalars, fermions and vectors; $f_b(n)$ was recently computed for free conformal scalars and fermions \cite{Lee:2014xwa}, and found to equal $f_c(n)$. It was conjectured in \cite{Lee:2014xwa} that, for general CFTs,
\be\la{fbfc}
f_b(n) = f_c(n)~.
\ee
We later provide new arguments as to when this may or may not be the case. Note that neither $f_b(n)$ nor $f_c(n)$ has ever been computed holographically. 

In sum, $f_a(n)$, $f_b(n)$ and $f_c(n)$ appear to be independent functions, although $f_b(n)=f_c(n)$ in a to-be-determined class of theories that includes free conformal scalars and fermions. 

\subsubsection{New results}
In fact, our perturbative results \eqr{flatgeo} and \eqr{fs} imply that $f_c(n)$ is not independent of $f_a(n)$. Starting from \eqr{fursaev}, consider a linearized geometric perturbation around flat space. As discussed in \cite{RS1}, only the Weyl term contributes at this order:
\be\la{cftfirst}
\d^{(1)}S_n = \left({f_c(n)\over 2\pi}\int_{\Sigma} C^{ab}_{~~ab}\right)\log R/\eps~.
\ee
We now compare this to our general perturbative expression \eqr{flatgeo} in $d=4$.
We are interested in extracting the log term that is linear in curvature fluctuations from the integral
\be\la{357}
\d^{(1)}S_n = {nF(n)\over 2(1-n)} \int_{\S} d^{2}y \int_0^{2\pi} d\t \int_{\eps/R}^{\infty} dr ~{1\over r^3}\left( {h^{\t}}_{\t}-{1\over 3}h^r_{~r} - {1\over 3}\sum_{i=1,2} {h^{y_i}}_{y_i}\right)~.
\ee
We have imposed a cutoff at $r=\eps/R$ near the entangling surface at $r=0$. By dimensional analysis, a $\log\eps/R$ term will only arise from terms of $O(r^2)\sim O(x^2)$. Examining (\ref{metricexp}), we only need the following terms of the induced metric fluctuation:
\begin{eqnarray}\la{hfirst}
h_{i j}=x^a x^c R_{i a c j}~,  & & h_{a b}=-\frac{1}{3} R_{a c b d} x^c x^d~.
\end{eqnarray}

Using $(x^1,x^2)=(r\cos\t,r\sin\t)$, one passes to polar coordinates in the transverse space and straightforwardly integrates \eqr{357}. The angular integrals require that $x^ax^c\rar \d^{ac}r^2$ times some angular functions. Extracting the logarithmic piece of the radial integral yields the following result for $\d^{(1)}S_n$:\footnote{We have used the result that under the integral over $\S$ without boundary,
\be
\int_{\Sigma}C^{ab}_{~~ab} = {1\over 3}\int_{\S}\left( \d^{ac}\d^{bd}R_{abcd} + \d^{ac}\d^{ij}R_{iacj}\right)
\ee
}
\be
\d^{(1)}S_n = \left({\pi\over 2} {n\over n-1}F(n)\int_{\Sigma} C^{ab}_{~~ab}\right)\log R/\eps ~.
\ee
Equating this with \eqr{cftfirst}, we obtain
\be\la{fcF}
f_c(n) = \pi^2{n\over n-1} F(n)~.
\ee
This gives a first principles definition of $f_c(n)$: it is proportional to the energy density on the conical spacetime $\C_n\times \IR^{2}$. Furthermore, we can rewrite \eqr{fcF} by trading $F(n)$ for $f_a(n)$ using \eqr{fs} and \eqr{sfa}, upon which we arrive at the following relation:
\be\la{fcfa}
f_c(n) = {n\over n-1} \big(a-f_a(n) - (n-1)f'_a(n)\big)~.
\ee
This is one of our main results. We have shown that if $\int_{\S}\Tr K^2-\frac{1}{2} K^2=0$, the $n$-dependence of 4d CFT vacuum R\'enyi entropy across an arbitrary surface $\S$ is fully fixed by $f_a(n)$.
\vs
In Section 4, we will discuss some applications of this result, including the first derivations of R\'enyi entropy at strong coupling across non-spherical surfaces. For now let us make some comments:\vs

\bul It is straightforward to check \eqr{fcfa} using free field results \cite{Casini:2010kt,Dowker:2010bu,Fursaev:2012mp}. It is also manifestly consistent with the known behavior of these functions at $n=1$, 
\be\la{fc'fa'}
f_c(1) = -2f'_a(1) = c~.
\ee
This is rather interesting. In Solodukhin's formula for EE -- the $n=1$ version of \eqr{fursaev} -- $a$ and $c$ appear as independent quantities inherited from the trace anomaly. \eqr{fc'fa'} trades these parameters for Taylor series coefficients of a {\it single} function near $n=1$. This speaks to the ability of R\'enyi entropy to recontextualize known results about EE and other properties of CFTs.

\bul  Our result has surprising implications for the correlators of modular Hamiltonians for non-spherical surfaces, despite their non-locality. In particular, the logarithmically divergent part of these correlators is largely fixed by correlators of the CFT stress tensor. The eager reader may skip to Section 6.2, where we discuss this in more detail.

 \bul Some examples of surfaces that turn on $f_c(n)$ are black hole horizons, as black hole backgrounds generically have a non vanishing Weyl tensor; see \cite{Solodukhin:2011gn} for a detailed overview.
 
\bul One can show, using \eqr{fcfa}, that the only pair $(f_a(n), f_c(n))$ for which $f_a(n)/f_c(n)$ is independent of $n$ is that of the free scalar \cite{Casini:2010kt},
\be
f_c(n) = 3f_a(n) = 3a{(1+n)(1+n^2)\over 4n^3}~.
\ee
This follows from \eqr{fcfa} with the boundary condition $f_a(n\rar 0)\sim n^{-3}$, which is the scaling of the thermal free energy of a CFT on $\IR^4$ \cite{Swingle:2013hga}.

\bul Using \eqr{fcF}, we can distill our earlier comments about energy density in cosmic string backgrounds to the following statement: in $d=4$, this energy density is proportional to the universal part of the R\'enyi entropy fixed by the background Weyl curvature. 
 
\bul A similar first order analysis was done in \cite{RS1} for EE rather than R\'enyi entropy. The authors checked the consistency of the perturbative approach against the $n=1$ version of the formula \eqr{fursaev}. Interestingly, the generalization to R\'enyi entropy affords new insights with no cost in technical difficulty.

\subsection{$d=6$}
The $d=6$ story is precisely analogous to the $d=4$ story. In $d=6$, the R\'enyi entropy functional takes the form \cite{Hung:2011xb}\cite{Safdi:2012sn}
\begin{equation} \label{univ6d}
S_n = \left(2 f_a(n) \int_{\Sigma} E_4 + 8\pi f_{B_3}(n) \int_{\S} \nabla^2 C^{ab}_{~~ab} + \ldots\right)\log R/\eps
\end{equation}
where the $\ldots$ represent terms that are at least quadratic in either the Weyl tensor or extrinsic curvatures. As such, $f_{B_3}(n)$ is the $d=6$ analog of $f_c(n)$, and clearly, $f_a(n)$ plays the same role here as in $d=4$. $a$ and $B_3$ are the Weyl anomaly coefficients \cite{Bastianelli:2000hi},
\begin{equation}\la{6dtrace}
\langle T^{\mu}_{~~ \mu} \rangle = \sum_{i=1}^3 B_i I_i + 2a E_6
\end{equation}
where $E_6$ is the $d=6$ Euler density, and the $I_i$ are independent invariants formed from the Weyl tensor. Our normalization is such that at $n=1$, $f_a(1) = a$ and $f_{B_3}(1) = B_3$ \cite{Hung:2011xb}. For $\S=S^4$, 
\be\la{sph6}
S_n|_{\rm log} = 4 f_a(n)\log R/\eps = {4\over \pi^2}f_a(n) \Vol(\IH^5)~.
\ee
Perturbing \eqr{univ6d} around $\Sigma = S^4$, the first order perturbation $\d^{(1)}S_n$ is
\begin{equation}\la{6dpertcft}
\delta^{(1)}S_n = 8\pi f_{B_3}(n) \int_{\S} \nabla^2 C^{ab}_{~~ab}\log R/\eps~.
\end{equation}
We wish to equate this to the logarithmically divergent part of our perturbative expression
\begin{equation}\la{6dpert}
\delta^{(1)}S_n = {n\over 2(1-n)} F(n) \int_{\S}d^4y \int^{2\pi}_0 d\t \int_{\eps}^{\infty} {dr} {1\over r^5}\left(h^{\t}_{~\t}-{1\over 5}h^r_{~r}-{1\over 5}\sum_{i=1}^4h^{y^i}_{~y^i}\right) ~.
\end{equation}
This time, a log term is generated by components of $h$ that are of $O(r^4)\sim O(x^4)$. In Appendix A, we compute the transverse components of \eqr{6dpert}; upon matching to \eqr{6dpertcft}, we obtain the desired relation,
\begin{equation}\la{fb3F}
{f_{B_3}(n)} =  {1\over 384}{n\over n-1}F(n)~.
\end{equation}
Trading $F(n)$ for $f_a(n)$ using \eqr{fs} and \eqr{sph6}, 
\be\la{fb3fa}
f_{B_3}(n) = \frac{1}{192 \pi ^3} \frac{n}{ n-1} \left( a-f_a(n)-(n-1) f_a'(n) \right)~.
\ee
Equations \eqr{fb3F} and \eqr{fb3fa} are the $d=6$ analogs of equations \eqr{fcF} and \eqr{fcfa}, respectively. 

\subsection{General even dimensions}

Recall the general formula \eqr{Renyischem}. It is clear that in all even $d$, a linearized perturbation around a spherical entangling surface will express the $d$-dimensional analog of $f_c(n)$ (the coefficient of the term linear in the Weyl tensor) in terms of $f_a(n)$ (the coefficient of the $(d-2)$-dimensional Euler density) and its first derivative. The $n$-dependence takes the form in \eqr{fcfa} and \eqr{fb3fa}.

In fact, the perturbative approach suggests a new way to think about this formula: given geometric perturbations of magnitude $\eps$ around a spherical entangling surface, \eqr{Renyischem} can be viewed as the truncation, at $O(\eps^{d-2})$, of the perturbative expansion of the universal part of the R\'enyi entropy. The key point is that we are only interested in the logarithmic divergence, and the number of terms which can possibly contribute is fixed by dimensional analysis. In $d$ dimensions, the $O(r^{d-2})$ term in the derivative expansion of the metric contributes linearly to the log, and can contribute at most $O(\eps^{d-2})$. Similarly, one expects a term of $O(r^{\frac{d-2}{2}})$ to contribute quadratically to the log; and so on. So in $d$ dimensions, we obtain the complete expression for the log term just by expanding the metric to $O(r^{d-2})$ and considering the perturbative corrections to the R\'enyi entropy of the sphere through $O(\eps^{d-2})$. Note that this argument follows from conformal invariance, which means that, even if {\it a priori} one must deal with up to $d-2$ integrals over the surface, one ends up with a single local integral.

In odd dimensions, the situation is different: as the universal term is a constant, to understand the shape dependence one must deal with the explicit geometries. The perturbative expansion will give an infinite number of terms that can contribute. Some studies of the shape dependence of the EE in odd dimensions have been done in \cite{Klebanov:2012yf,Allais:2014x}. We will return to the results of \cite{Allais:2014x} in the next section.		

\section{Applications}
We discuss various applications of our work so far. 

\subsection{$f_c(n)$ at strong coupling}
Equation \eqr{fcfa} allows us to compute $f_c(n)$ at strong coupling. As we will argue in Section 5, it seems that for CFTs dual to 5d Einstein gravity, $f_b(n)=f_c(n)$. For any theory in which this is true, the following results give the R\'enyi entropy across the cylinder as well.

The R\'enyi entropy across the sphere was computed in \cite{Hung:2011nu} for a family of gravitational theories with CFT duals. As we have seen in \eqr{twist} and \eqr{fcF}, we can write $f_c(n)$ (and its higher dimensional analogs) simply in terms of the dimension of the spherical twist operator, $h_n$. For simplicity, consider CFTs dual to pure $(d+1)$-dimensional Einstein gravity. Temporarily using $f_c(n)$ to denote the coefficient of the term linear in the Weyl tensor in any even-dimensional CFT, the ratio $f_c(n)/f_c(1)$ at strong coupling is given by
\be\la{fcratio}
\frac{f_c(n)}{f_c(1)}=\frac{1}{n-1} \frac{h_n}{h'_{1}}={d-1\over 2}\left(\frac{n}{n-1} x_n^{d-2}(1-x_n^2)\right)~,
\ee
where
\be
x_n=\frac{1}{n d} (1+\sqrt{1-2 d n^2+d^2 n^2})~.
\ee
The prime denotes a derivative with respect to $n$. These results can be easily extended to Gauss-Bonnet and quasi-topological gravity using the results of \cite{Hung:2011nu}.

For $d=4$, one can go even further and compute the leading $O(\lambda^{-3/2})$ and $O(\lambda^{1/2}N^{-2})$ corrections to $f_c(n)$ in ${\cal N}=4$ super-Yang-Mills. These appear as $O(\a'^3)$ and $O(\a'^3g_s^2)$ corrections, respectively, to hyperbolic black hole entropy in type IIB supergravity \cite{Galante:2013wta}. Using the conventions of \cite{Galante:2013wta}, $f_c(n)$ is given by
\es{}{
f_c(n)&=\frac{3\pi^2 (1+n)}{2 n^3} \left ( \frac{L}{\ell_p} \right )^3 \\ &\times\left ( \frac{1}{32} \frac{(1+\sqrt{1+8 n^2})^3}{3+\sqrt{1+8 n^2}} +10 \gamma \frac{8 n^2-3+9\sqrt{1+8 n^2}}{\sqrt{1+8 n^2}(3+\sqrt{1+8 n^2})^3} (1-n^2)^2+O(\a'^4)\right )
}
with $\gamma=\frac{1}{8} \zeta(3) \alpha'^3/L^{6}(1+ {\pi^2\over 3\zeta(3)}g_s^2)$,  plus non-perturbative corrections \cite{Green:1997tv}. In gauge theory variables, $\gamma = {1\over 8}\zeta(3)/\lambda^{3/2}(1+ {1\over 48\zeta(3)}\l^2/N^2)$
and $(L/\ell_P)^3 = (N/2\pi)^2$. 

These constitute the first computations of R\'enyi entropy across non-spherical entangling surfaces in any strongly coupled CFT.

\subsection{Entanglement negativity}
The logarithmic entanglement negativity (henceforth, negativity), $\E$, is a measure of the amount of distillable entanglement present in a given state \cite{Vidal:2002zz,Calabrese:2012ew,Calabrese:2012nk}. This is to be distinguished from ordinary EE, which is also sensitive to bound entanglement. For a bipartite system,
\be\la{neg}
\E = S_{1/2}
\ee
where the right-hand side is the ordinary R\'enyi entropy at $n=1/2$. 

In \cite{Rangamani:2014ywa}, negativity across spherical entangling surfaces in flat space CFTs at both weak and strong coupling was studied using the relation \eqr{neg}. In particular, \cite{Rangamani:2014ywa} suggested that the ratio of (the universal parts of) $\E$ to ordinary EE in $d$ dimensions,
\be
\X_d = \left|{S_{1/2,\rm univ}\over S_{1,\rm univ}}\right|
\ee
may play a privileged role in the relation between spacetime and entanglement. $\X_d$ was conjectured to always be greater than or equal to one, which was shown in various examples for spherical bipartitions. 

Our relations \eqr{fcfa} and \eqr{fb3fa}, evaluated at $n=1/2$, thus permit immediate evaluation of $\X_d$ for a class of non-spherical bipartitions in 4d and 6d CFTs, given knowledge of $f_a(n)$. The ratio \eqr{fcratio} evaluated at $n=1/2$ is precisely $\X_d(\S_c)$ evaluated for even-dimensional holographic CFTs dual to Einstein gravity, where $\S_c$ is a hypothetical entangling surface that isolates the $f_c(n)$ term.\footnote{Clearly, our results also permit evaluation of $\X_d$ for any surface that turns on a linear combination of the $f_c(n)$ and $f_a(n)$ terms.} Denoting this holographic ratio as $\X_d^{\rm hol}(\S_c)$, one can easily show that $\X_d^{\rm hol}(\S_c)$ is a monotonically increasing function of $d$, and that $\X_d^{\rm hol}(\S_c)=1$ at the low value $d\approx 1.599$ where this ratio ceases to make sense anyhow. In $d=4,6$, for example, we find 
\es{}{
\X_4^{\rm hol}(\S_c) &\approx 2.424~,\\
\X_6^{\rm hol}(\S_c) &\approx 2.600 ~.
}
It is perhaps worth noting the peculiar result that in the large $d$ limit, $\X_{\infty}(\S_c)= e$. This is reminiscent of similar behavior for the spherical case studied in \cite{Rangamani:2014ywa}.

One can also use the results of \cite{Fursaev:2012mp} to show that $\X_4^{\rm hol}(\S_c)\approx 0.615 \X_4^{\rm free}(\S_c)$. This is the same behavior that was observed in \cite{Rangamani:2014ywa}. 

In sum, these calculations using non-spherical surfaces lend further support to the conjecture that $\X_d\geq 1$ for general bipartitions.


\subsection{Geometric perturbation around non-vacuum states}

It is hopefully clear from the analysis of Section 2 that we can also apply our results for small deformations around non-vacuum states. Because there is no analog of Solodukhin's formula for  states other than the vacuum, these applications are necessarily perturbative in the deformations.\footnote{ Similarly, our perturbative analysis applies straightforwardly to the charged R\'enyi entropies of \cite{Belin:2013uta}.}

As an example, we will show how this works in the background of a Wilson loop. The EE of the Wilson loop has been considered in \cite{Jensen:2013ora,Lewkowycz:2013laa}.  Here we will see how the R\'enyi  entropy of a Wilson line in a disk changes as we slightly perturb the background geometry. We simply apply our formula \eqr{fogeo} in the presence of the Wilson loop, $W$:
 \be
 \delta^{(1)} S_n(W)=\frac{n}{2(1-n)}\int\big( \langle T_{\mu \nu} W \rangle_n - \langle T_{\mu \nu} W \rangle_1\big) h^{\mu\nu}~.
\ee
In this case, there will also be a geometrical factor that factorizes because of conformal invariance, so in the notation of \cite{Lewkowycz:2013laa}, $\langle T_{\mu \nu} W \rangle_n=\frac{h_w(n)}{h_w(1)} \langle T_{\mu \nu} W \rangle_1$, where $\langle T_{\mu \nu} W \rangle_1$ is given in \cite{Lewkowycz:2013laa} and $h_w(n)$ is a theory-dependent function of $n$.  So we can use the previous techniques to compute $\langle T_{\tau \tau} W \rangle_n$ and thus determine $h_w(n)$.

The R\'enyi entropy for the Wilson loop is not known in general. However, it is known at strong coupling, where it is given by the contribution to the entropy of a hyperbolic black hole from a string. For a black hole with horizon curvature scale $R$, the R\'enyi entropy for the sphere at strong coupling is 
\begin{equation}\label{swn}
S_n(W)=\frac{R^2}{\alpha' } \frac{1}{d(1-n)} \left(\sqrt{(d-1)^2 n^2+1-n^2}+1-d n\right)~.
\end{equation}
When $n=1$, we get the usual $S_{EE}=\frac{R^2}{\alpha'(d-1)}$. 
We can apply our formula there and the R\'enyi entropy for a small deformation is:

\begin{equation}
\delta^{(1)} S_n(W)= \frac{1}{2\pi \text{Vol}(S^{d-2})}\frac{n}{2(n-1)} \left [ 1-\left (1+\frac{1-n^2}{(d-1)^2 n^2} \right )^{-1/2} \right ] \int_{\M} \langle T_{\mu \nu} W \rangle_{1}  h^{\mu \nu}~.
\end{equation}
To get the final answer, one should insert an explicit expression for $h^{\mu\nu}$. The expression for $\langle T_{\mu \nu} W \rangle_{1}$ can be found in \cite{Lewkowycz:2013laa}.

\subsection{$n$ limits}

Consider the $n\rar 0$ limit of $f_a(n)$. From the hyperboloid perspective, $\beta=2 \pi n$, so the $n \rightarrow 0$ limit corresponds to the high temperature regime in flat space \cite{Swingle:2013hga}.  The R\'enyi entropy becomes $S_0=\log Z_0\sim n^{-(d-1)}$. In $d=4$, from \eqr{fcfa} we obtain $f_c(0)=3 f_a(0)$, as follows from the relation between free energy and energy in flat space.

The $n \rightarrow \infty$ limit is equivalent to computing the partition function of the hyperboloid at zero temperature. In this limit, $S_{\infty}= -\frac{1}{n} \log Z_n+\log Z_1=F_{\infty}-F_{1}$. Now, $F_{\infty}$ is the free energy of the hyperboloid at zero temperature, that is, the Casimir energy of the CFT on $S^1\times \IH^{d-1}$. So, at large $n$, the $f_i(n)$ asymptote to constants: $f_c(\infty)=a-f_a(\infty)$.

\subsection{Shape dependence in odd-dimensional CFTs}

It was recently shown in \cite{Allais:2014x} that in any $d$-dimensional CFT in a {\it fixed} flat background, the geometric integrals appearing in our linearized analysis around the sphere are only sensitive to the breathing mode of the sphere. Using this result,\footnote{This observation, along with essential details, also appears in \cite{Allais:2014x}. We thank Mark Mezei for discussions leading to this result.} one can show that the linearized change in the universal term of the R\'enyi entropy vanishes in all $d$. This is non-trivial in odd $d$ in particular, where the R\'enyi entropy is non-local, reflected in the absence of formulae like \eqr{Renyischem}.

\section{Second order perturbations and the relation between $f_b(n)$ and $f_c(n)$}

We now move onto second order geometric perturbations. For concreteness, we continue to study perturbations of 4d CFT vacuum R\'enyi entropy across a sphere in flat space. From \eqr{fursaev}, it is clear that second-order perturbations isolate the $f_b(n)$ term, which is quadratic in extrinsic curvature. Thus, we can obtain a definition of $f_b(n)$ by equating that term with the $\Oc(K^2)$ contribution to our perturbative expressions \eqr{sum}, with $O=\half T^{\mu\nu}h_{\mu\nu}$: that is, 
\begin{equation}\label{51}
\delta S_n|_{K^2}=\delta^{(1)} S_{n}|_{K^2}+\delta^{(2)} S_{n}|_{K^2}=-\frac{f_b(n)}{2\pi} \int_{\Sigma} \big( {K^a_{i j}} {K^a_{j i}} -\frac{1}{2} (K^a_{ i i})^2\big)\log R/\eps
\end{equation}
defines the function $f_b(n)$. 

The first term is first-order in perturbation theory from a second-order deformation of the metric. So we evaluate \eqr{357}, now with (cf. \eqr{metricexp})
\be
h_{ij} = x^ax^cK_{cil}K_{a~j}^l~.
\ee
The contribution from this term is equivalent to the contribution from $h_{ij}$ in \eqr{hfirst}, with $R_{iacj} \rar K_{cil}K_{a~j}^l$. Using that equation and \eqr{fcF}, we obtain
\begin{equation}
\delta^{(1)} S_{n}|_{K^2} = \left({f_c(n)\over 6\pi}\int_{\S}{K^a_{i j}} {K^a_{j i}}\right)\log R/\eps~. \label{metric2KK}
\end{equation}
Already, we see that $f_c(n)$ appears in the definition of $f_b(n)$. 

We turn to the remaining contribution $\delta^{(2)} S_{n}|_{K^2}$,
\bea\la{54}
\delta^{(2)} S_n|_{K^2} =\frac{1}{8(1-n)} \int_{\C_n \times \IR^2} d^4z' \int_{\C_n \times \IR^2} d^4z \langle T^{ij}(z') T^{kl}(z) \rangle_n h_{ij}(z') h_{kl}(z) 
\eea
with
\be\la{55}
 h_{ij}=2 K_{ij}^a x^a~.
 \ee
We have suppressed coordinate dependence of the extrinsic curvatures.\footnote{We have defined $\lbrace z^{\mu}\rbrace=\lbrace x^a,y^i\rbrace$, where $a=1,2$, and $i=1\ldots d-2$ as in previous sections. In the next subsection, we will also sometimes use $x^{\mu}$ to denote the full set of coordinates.} We have not written the terms $\int \int \langle T T \rangle_1 h h$ because these terms do not see the replicated geometry. These terms also appear in the expansion of the partition function and, after we regularize them properly,\footnote{For example one can express the two-point function as derivatives acting on $\frac{1}{(x-y)^4}$ and integrate it by parts. If $h$ is smooth, it won't diverge.} they will be finite. 

It is difficult to extract the logarithmic divergence from \eqr{54}. We will not reach the end of the calculation, but would like to sketch a strategy that we believe underlies the final answer. Let us explain why this is a difficult computation by way of a primer on general aspects of CFT on the cone.

\subsection{Conformal symmetry and the cone}

Given a Lorentz invariant CFT in flat space, one can fully constrain the structures that appear in two- and three-point functions  \cite{Osborn:1993cr,Erdmenger:1996yc}. There is no conformal invariant scalar; the only vector that can appear is $\hat{x}_{\mu}=\dfrac{x_{\mu}}{|x|}$; the only two-tensors that can appear are $\delta_{\mu \nu}, \hat{x}_{\mu} \hat{x}_{\nu}$; and so on. Because we have translation invariance, $\langle O(x) \rangle=\langle O(0) \rangle=0 $. Imposing tracelessness and conservation constrains the number of possible structures that can appear in the stress tensor correlators.  There is one possible structure for the stress tensor two-point function and three for the three-point function. Because there is no conformal invariant scalar, these correlators will then be determined by one ($C_T$) and three (${\cal A},{\cal B},{\cal C}$) constants, respectively.\footnote{A Ward identity fixes one linear combination of (${\cal A},{\cal B},{\cal C}$) in terms of $C_T$ \cite{Osborn:1993cr}.}

Now moving to the cone, with metric $ds^2=dr^2+r^2 d\tau^2+\delta_{ij}dy^idy^j$ and $\tau \sim \tau+2 \pi n$, we are introducing a new vector to the game: $\xi=r \partial_{\tau}$. This vector breaks translation invariance in the $(x_1,x_2)$ directions, here written in polar coordinates. In general, we don't expect the theory in the replicated geometry to break any global symmetry. \footnote{There are some particular cases in settings with a large number of degrees of freedom where for $n$ large enough there is spontaneous breaking of some global continuous symmetries \cite{Belin:2013dva}.} The number of structures that can appear is now much greater, and we have a conformal invariant scalar: $\hat{r}=\frac{x^{\mu} \xi_{\mu}}{|x|^2}=\frac{r^2}{r^2+y^2}$. There is also a new number, $n$, so even if we could constrain higher-point functions in a similar way as in flat space, the structures could be multiplied by an arbitrary function of $n$ and $\hat{r}$. This makes conformal symmetry much less restrictive.

Due to the translation breaking in the $(x_1,x_2)$ directions, one-point functions need not vanish. 
But they are constrained nonetheless. For scalar operators $O$,
\be\la{onept}
\langle O(r,\tau,y) \rangle_n= \langle O(r,0,0) \rangle_n=\frac{a_{O}(n)}{r^{\Delta_O}}
\ee
where $a_O(n)$ is some function of $n$. The one-point function of a vector is fixed to be zero by parity: the only vectors that can appear are $\xi$ or $\hat{x}$, but they are not symmetric with respect to $\tau \rightarrow -\tau$, so the expectation value must be zero if parity is not spontaneously broken.\footnote{Note that this is not true if there is a background magnetic field that breaks parity explicitly. See for example \cite{Belin:2013uta}.} For the stress tensor, symmetries fix $\langle {T^{\mu}}_ {\nu} \rangle_n=-\frac{F(n)}{(d-1)r^d} ({\delta^{\mu}}_{\nu}-d \xi^{\mu} \xi_{\nu})$. The two-point function will be much less constrained, which is the crux of the problem in computing \eqr{54}. A classification of the possible structures that could appear is a challenging undertaking that we do not pursue here. 

We expect that one should be able to apply familiar CFT techniques on the cone. In particular, one can perform an operator product expansion (OPE). The OPE is an statement about how two operators collide inside correlators. In computing two-point functions, then, the only difference compared to flat space CFT is that now one-point functions may be nonzero. More explicitly, the two-point function now takes the form
\es{}{
\langle O_1(z') O_2(z) \rangle_n &= \sum_{O} c_{1 2 O} |z'-z|^{\Delta_{O}-\Delta_1-\Delta_2} \langle  O(z) \rangle_n\\&=\sum_{O} c_{1 2 O} |z'-z|^{\Delta_{O}-\Delta_1-\Delta_2} \frac{a_{O}(n)}{r_z^{\Delta_O}} ~.
}
We have suppressed operator indices for clarity. Note that in general, $c_{12 O}=c_{12 O}(\hat{z'}-\hat{z})$.

\vs
\bul {\bf Conformal maps}
\vs

Let us also point out that when one performs a conformal transformation, $r \rightarrow l_{c}$, where $l_c$ is the conformal distance to the conical deficit/entangling surface. This means that one-point functions can be evaluated in the easiest conformal frame, and transformed by the above replacement (up to the conformal anomaly). Some examples of geometries and their corresponding $l_c$ are:
\begin{eqnarray*}
\text{Cone}:&  dr^2+r^2 d\tau^2+\delta_{ij}dy^idy^j  &l_c=r \\
\text{Branched sphere}: & d\theta^2+\sin^2 \theta d\tau^2+\cos^2 \theta d\Omega_{d-2}^2 & l_c=\sin \theta \\ 
\text{Hyperboloid}: & d \tau^2+du^2+\sinh^2 u d\Omega_{d-2}^2  & l_c=1 \\ 
\text{Spherical entangling surface}: & dt^2+dr^2+r^2 d\Omega_{d-2}^2 & l^2_c=\dfrac{(r^2+t^2-R^2)^2+4 R^2 t^2}{4 R^2}
\end{eqnarray*} 
These geometries are all related to each other by straightforward conformal transformations. 

\subsection{An OPE argument}

We now sketch a method to extract the logarithmically divergent term in \eqr{54}. As we discussed in the previous section, this integral seems hard to tackle only using symmetries. We would like to pursue a different direction: in particular, we will provide some arguments for how to ``fish" the $\log \epsilon$ piece, by using the OPE. These arguments are rather schematic, but we believe that a formal version of the following will provide the proper definition of $f_b(n)$. 

We focus on the stress tensor two-point function, $\langle T^{ij}(z') T^{kl}(z) \rangle_n$. 
Because we are integrating over all space, the OPE may seem irrelevant. However, the logarithmically divergent contribution to the integral should come from the neighborhood of the entangling surface which sits at the tip of the cone. This region is described by the limit $x \sim 0, x' \sim 0$, but finite $(y,y')$. In this limit, the OPE is an expansion in $1/r$. A word of caution is in order: because we are integrating the space surrounding the conical deficit, there could be kinematic configurations for which the OPE does not seem to converge.\footnote{For example, naively it seems that if we have two operators on opposite sides of the singularity, the radius of convergence of the OPE does not extend between the operators. However, from the hyperboloid perspective, these two points are separated a finite distance along the $S^1$ and are far away from the singularity at the boundary. We thank Alexander Zhiboedov for discussions on this matter} We are going to ignore this issue in our discussion.

Substituting the $TT$ OPE into \eqr{54}, we can shift coordinates $z' =z+\delta z$. The resulting integral looks as follows:
\es{int2}{
\delta^{(2)} S_n|_{K^2} &=\frac{1}{2(1-n)} \sum_{O}a_{O}(n)\int d^2y \int d^4\d z K^a_{ij}(y+\d y)K^b_{kl}(y)c_{TTO}(\d \hat z)|\d z|^{\D_O-8}\\
&\times\int d^2 x (x^a+\d x^a)x^b{1\over r_z^{\D_O}}
}
where $r_z^2 = x_1^2 + x_2^2$. All integrals are over the full conical spacetime $\C_n\times \IR^2$. Our job is to extract a logarithmic divergence from the integrals over $\d z$ and $x$. In particular, the physical divergence of the R\'enyi entropy should come from the integral on the second line: this is the integral that ``sees'' the singularity at $r=0$. 

Focusing now on the second line, the $x^b\d x^a$ term vanishes when integrated over $\t$, because it is odd in $x$. The integral over $x^ax^b$, on the other hand, only yields a log divergence for $\D_O=4$. 

The remaining integral over $\d z$ seems UV divergent and highly dependent on the details. But because it is a $z \rightarrow z'$  divergence, we expect that, after adding the proper contact terms, it should simply end up contributing as an order one constant, times some tensor structure. We emphasize that we don't expect any extra $n$-dependence from this integral, as it will be dominated by regularizing the UV data and thus will not see the covering space.

Thus, we have reached an interesting, if tentative, conclusion: only $\D=4$ operators which appear in the $TT$ OPE and have non-vanishing one-point function in the conical spacetime ${\cal C}_n\times \IR^{d-2}$ seem to contribute to the logarithmic divergence of the second order perturbations of spherical R\'enyi entropy. Denoting the $\D=4$ operators as $O_4$, and the result of the $\d z$ integral in the first line of \eqr{int2} as ${\cal T}^{ijkl}_{O_4}$, we can write the result of the previous manipulations schematically as
\es{int4}{\d^{(2)}S_n|_{K^2} = {n\pi\over 2(1-n)}\left(\sum_{O_{4}}a_{O_4}(n) \int d^2y K^a_{ij}K^a_{kl} {\cal T}^{ijkl}_{O_4}\right)\log R/\eps~.
}

In Appendix C, we discuss the structure of the sum \eqr{int4}. We are led to argue that $f_b(n)=f_c(n)$ in theories which satisfy the following property: they do not contain exactly marginal scalar operators or conserved spin-2 currents besides the stress tensor which appear in the $TT$ OPE {\it and} have nonzero expectation value in $\C_n\times \IR^2$. Said another way, consideration of the stress tensor contribution alone to \eqr{int4} seems to imply $f_b(n)=f_c(n)$, but if there are other operators $O_4$ that also contribute, these may modify the relation between $f_b(n)$ and $f_c(n)$. We argue that the class of theories for which $f_b(n)=f_c(n)$ includes the following: \vs

a) Free scalars, fermions and Maxwell fields\vs

b) $\N=4$ SYM \vs

c) Strongly coupled CFTs dual to classical gravitational theories with minimally coupled scalar fields, and/or a family of hyperbolic black hole solutions without massless scalar hair.

\vs

This includes all Einstein gravity compactifications of type IIB supergravity on compact 5-manifolds. Again, the details can be found in Appendix C. 
\vs
The presentation of these arguments has been quite schematic. A more in-depth treatment must address the issue of OPE convergence in the conical background, and the details of the integrals that lead to the EE result $f_b(1)=f_c(1)=c$. We leave this interesting pursuit for the future.

\section{Are there more direct methods to compute the R\'enyi entropy?}
We now discuss some implications of our results for the structure of a generic modular Hamiltonian, and address the section title for generic shapes. 

\subsection{The rarity of local modular Hamiltonians}

Modular Hamiltonians are generically complicated and non-local. For entanglement across the plane, being the same thing as thermal entropy of Rindler space, the modular Hamiltonian is an integral of the stress tensor. If our theory has conformal invariance, one can perform a conformal transformation of the half-plane to a sphere and the modular Hamiltonian of the sphere will still be given by the integral of a local stress tensor. These modular Hamiltonians are local in $\tau$, the coordinate conjugate to the replica parameter $n$. Locality means that the operator that performs the replica trick $\int_{S^1 \times M} T_{\tau \tau}(\t)=2\pi \int_{M} T_{\tau \tau}$ does not depend on $\tau$.\footnote{This follows from conservation of $T_{\tau \tau}$.}

Now we would like to argue that, at least for CFTs in flat space, the sphere and its conformal counterparts are the {\it only} surfaces whose modular Hamiltonian will be the integral of a local operator. Let us specify to $d=4$. Recall that in any dimension, the $m$'th derivative of the R\'enyi entropy at $n=1$ is proportional to a connected $(m+1)$-point function of $K$, cf. \eqr{correxp}. The first derivative behaves as $S'_{n=1} \propto \langle K K \rangle_{1,c} $. This contains a logarithmic term, which we compare to a derivative at $n=1$ of the formula \eqr{fursaev}. Now, if $K$ is a local integral of a stress tensor, $S'_{n=1} \propto c$. But this only holds if the entangling surface only turns on $f_a(n)$, as for the sphere. That is, $f_b'(1) \not\propto c$. One might wonder whether there are non-spherical surfaces with a vanishing extrinsic curvature term, but, as was shown in \cite{Astaneh:2014uba}, in flat space there are no such surfaces. So we conclude that, in flat space, the only local modular Hamiltonian is that of the sphere. 

In other words, there doesn't exist a complicated conformal transformation that would make local the modular Hamiltonian of an entangling surface such as the cylinder or the stripe. In the next subsection, we will present a candidate geometry, without local modular Hamiltonian, whose partition function we expect to give us the R\'enyi entropy. 

The above argument was in $d=4$, but we expect it to generalize to all dimensions. In addition, we have not ruled out the possibility that, if the background Weyl tensor is nonzero, there are non-spherical surfaces that only turn on $f_a(n)$. But these surfaces should be regarded as in the same equivalence class of R\'enyi entropy as the sphere, as their entanglement spectra have identical dependence on $n$.

\subsection{Correlators of generic modular Hamiltonians}
If modular Hamiltonians are generally non-local, what else useful can be said? Let us return to our result \eqr{fcfa} relating $f_a(n)$ and $f_c(n)$ in $d=4$. The perturbative expansion of \eqr{fcfa} around $n=1$ implies the following. Consider the modular Hamltonian associated to a generic surface in a background with nonzero Weyl curvature. Then the logarithmic part of its $m$-point correlators are fixed by $m$- and $(m+1)$-point correlators of the local modular Hamiltonian for the sphere, $K_{\rm Sph}$, up to extrinsic curvature contributions. 

To demonstrate this, we take the first derivative of $f_c(n)$ at $n=1$, and use \eqr{sfa} and \eqr{correxp} to trade $n$-derivatives for correlators of modular Hamiltonians. Then for {\it any} entangling surface $\S$ with modular Hamiltonian $K$, we have the following relation:
\begin{eqnarray}\la{KK}
&&\half \lang K K \rang_{1,c}\big|_{\log} \nonumber\\&=& \left[{\lang K_{\rm Sph}K_{\rm Sph}\rang_{1,c} \over 16\pi}\int_{\S} R_{\S}+\frac{\lang K_{\rm Sph}K_{\rm Sph}\rang_{1,c}-\frac{1}{2} \lang K_{\rm Sph}K_{\rm Sph}K_{\rm Sph} \rang_{1,c}}{8\pi} \int_{\Sigma} C^{a b}_{~~ a b}\right]\Bigg|_{\log R/\eps} \nonumber \\&+&{f_b'(1)\over 2\pi}\int_{\Sigma} \big( {K^a_{i j}} {K^a_{j i}} -\frac{1}{2} (K^a_{ i i})^2\big)\log R/\eps~.
\end{eqnarray}
As we explained in Section 5, in at least some theories, $f_b(n)=f_c(n)$. In such cases, the last term in \eqr{KK} combines with the Weyl term, and the logarithmically divergent part of $K$ correlators is {\it fully} determined by $K_{\rm Sph}$ correlators.

This perturbative data about non-local modular Hamiltonians is a surprising feature of \eqr{fcfa} that deserves closer study.

\subsection{Hyperbolic geometries for non-spherical entangling surfaces}

Here we would like to just mention that one can use the transformation of \cite{ch2} to get a ``deformed hyperboloid'' geometry in the spirit of \cite{Lewkowycz:2013nqa}. Deformed (or squashed \cite{Fursaev:2013fta}), means that the expansion near infinity (close to the entangling surface) will depend on the coordinate $\tau$.  

As shown in \cite{ch2}, a simple conformal transformation maps flat space $\IR^d$ to the hyperboloid $\cH^d$:
\es{}{
 ds_{\mathbb{R}^d}^2 &= dt^2 + dr^2 + r^2 d\Omega_{d-2}\\
 &= \Omega^2 (d\tau^2+du^2+\sinh^2 u d\Omega_{d-2})
 }
 with $\Omega=(\cosh u + \cos \tau)^{-1}$ and coordinates related by
 \be
 t = {\sin \tau \over \cosh u + \cos \tau} \,, \qquad r = {\sinh u \over \cosh u + \cos \tau} ~.
 \ee
Now, the point is that one can use this same transformation to compute entanglement across a cylinder or stripe, just by rewriting the transverse space $r^2 d\Omega_{d-2}$. For example, one can write the metric on $\IR^d$ in cylindrical coordinates:\footnote{ See also \cite{Guo:2013jva} for a similar application of the conformal transformation. }
\begin{equation}
ds^2_{\IR^d}=dt^2+dr^2+r^2 d\Omega_{d-2-m}+dy_{m}^2
\end{equation}
where $dy_m^2$ parameterizes an $\IR^m$ submanifold. A constant $t,r$ surface will be a $m$-cylinder, a surface with topology $\mathbb{R}^m \times S^{d-2-m}$: this is the topology of the entangling surface. The stripe is given by setting $m=d-2$ and taking $-\infty<r<\infty$. The previous conformal transformation will now map the unit cylinder to the boundary of a {\it deformed hyperboloid}. This has line element
\begin{equation}
ds^2= d\tau^2+du^2+\sinh^2 u d\Omega_{d-2-m}+(\cosh u+\cos  \tau)^2 dy_{m}^2~. \label{eq:cylhyp}
\end{equation}
 The boundary of the (Lorentzian) hyperboloid is mapped to the causal domain of the entangling surface: $u \rightarrow \infty, (r,t) \rightarrow (1,0);i \tau \rightarrow \pm   \infty ,(r,i t) \rightarrow (0,\pm 1)$. In other words, what this conformal transformation does is to map flat space to a deformed hyperboloid that covers only the causal domain of the $m$-cylinder. 

In the case of a sphere ($m=0$), a conformal mapping to the half plane renders $\tau$ the Rindler time coordinate. In the case of the cylinder, $\tau$ would become a coordinate that will look like Rindler time near the entangling surface, but will differ away from it because of extrinsic curvatures. 

The deformed hyperboloid is similar in spirit to the locally hyperbolic geometries of \cite{Lewkowycz:2013nqa} and, near $u=\infty$, it has the behaviour expected there. Now, the replicated geometry has to preserve the $\mathbb{Z}_n$ replica symmetry. This means that if we change the periodicity to  $\tau \sim \tau+ 2 \pi n$, the metric stays the same.

Near asymptotic infinity, the geometry becomes
\es{}{
 ds^2& = d\tau^2+du^2+\frac{e^{2u}}{4} d\Omega_{d-2-m}+ \frac{e^{2u}}{4}(1+2 e^{-u} \cos  \tau)dy^2_m+...\\& =\frac{1}{ r^2} \Big ( r^2  d\tau^2+dr^2+d\Omega_{d-2-m}+(1+2 r \cos  \tau)dy_m^2 \Big )+...
}
where $r=2 e^{-u}$. Stripping off the conformal factor, this is just the metric of a singular cone with extrinsic curvature.\footnote{We can conformally map this metric to that of the $m$-cylinder by conformally transforming the first sphere to a plane and the second plane to a sphere.} In even $d$, one can compute EE from the singular cone by extracting the contributions to the trace anomaly from the surface at $r=0$, as in \cite{Fursaev:2013fta}. This method involves regularizing the tip of the cone by a function $f(r,a)$ that interpolates between $n^{-2} r^{2(1/n-1)} $ near $r=0$ and unity at $r=\infty$.\footnote{This was the method used in \cite{Fursaev:2013fta}. Note that $r_{\rm there}=nr^{1/n}$.} Thus, the deformed hyperboloid can be used to extract the correct logarithmic terms in the EE.

We suspect that, as in \cite{ch2}, these geometries can be used to compute away from $n=1$. In particular, it is natural to suggest that the free energy of a CFT living in the deformed hyperboloid \eqr{eq:cylhyp} can be used to compute the R\'enyi entropy across the $m$-cylinder. This seems like a challenging computation; perhaps a derivative expansion around $n=1$ is feasible.\footnote{From this point of view, it seems that the modular Hamiltonian and the integral of the geometric (non-conserved) $T_{\tau \tau}$ may be related, even for these non-spherical entangling surfaces.} Also, it seems possible that this can be helpful in understanding the R\'enyi entropies of the $m$-cylinder from the gravitational perspective. One should be able to replicate the $n=1$ solution at least pertubatively in $n-1$. 

We leave a full investigation into the utility of the geometries (\ref{eq:cylhyp}) for the future. 

\section{Discussion}

In this work, we have used a general perturbative expansion of the R\'enyi entropy to make non-perturbative statements about its geometric dependence. By deforming the replicated geometry, we have been able to show that the function $f_c(n)$ is proportional to the one-point function of the stress tensor on the cone, and is a linear function of $f_a(n)$. 

We have also given some arguments about whether $f_b(n)=f_c(n)$ for general 4d CFTs. We have argued that it seems that only operators of dimension $\Delta=4$ can contribute to the derivation of $f_b(n)$ from the perturbative expansion, and in particular, the contribution of the stress tensor alone gives $f_b(n)=f_c(n)$. For the free theories, ${\cal N}=4$ SYM, and a wide class of holographic CFTs, we have argued that there is no contribution from other $\Delta=4$ operators and thus in these theories, $f_b(n)=f_c(n)$.

Of course, it is clear that the calculation of $f_b(n)$ and, more generally, the second order perturbation theory, needs further study. To do that one should be more careful and do a more honest calculation.

Aside from obtaining a more thorough understanding of the second order perturbation theory, there may be yet other ways to understand the underlying structure of R\'enyi entropy. For instance, we expect that some analog of the trace anomaly appropriate to CFTs on the cone at finite $n$ may exist that could be used to derive the R\'enyi entropy, analogous in spirit to the trace anomaly derivation of the EE functional. 

It is worth understanding what the relations we discovered among the $f_i(n)$ imply about the heat kernel approach to computing them. It would also be interesting to see if there is some easy way to understand the $n$-dependence when considering massive deformations \cite{Hertzberg:2010uv,Lewkowycz:2012qr}, maybe using the methods of \cite{RS3}. One might also wonder whether R\'enyi entropies in supersymmetric theories have simpler $n$-dependence or other hidden structure, much as their anomaly coefficients obey extra constraints. To see this, one may need to study super-R\'enyi entropy instead \cite{Nishioka:2013haa}.

Recall our derivation of the fact that the only entangling surface in $d=4$ flat space with a local modular Hamiltonian is the sphere. The simple math behind this statement -- namely, that the extrinsic curvature term in \eqr{fursaev} is a perfect square -- also suffices to show that the sphere minimizes the universal contribution of the EE among all possible surfaces \cite{Astaneh:2014uba}. The same statement also holds for R\'enyi entropy. It would be quite interesting if locality of the modular Hamiltonian and minimization of entanglement could be shown to be directly related.

Note that \eqr{renyifirstpi} is the R\'enyi equivalent of the first law of entanglement, which has been an active field of research in the context of holography \cite{Blanco:2013joa,Lashkari:2013koa,Faulkner:2013ica}. It would be nice to explore if its R\'enyi generalization can yield any further insight.

Let us zoom out and return to the philosophical question we posed in the introduction: why are we interested in the R\'enyi entropies? What do they give us that entanglement entropy doesn't? The R\'enyi entropies are less universal and thus more interesting in nature. For the sphere R\'enyi entropy, we know how to extract information from the CFT from its derivatives, since they contain information about all correlators of stress tensors. 

It is not very clear at the moment to what extent the shape dependence will give us more information about quantum and conformal field theories, or if shape dependence itself is too constrained. Each of these would be an interesting outcome. But, for example, a better understanding of the shape dependence of R\'enyi entropies in gravitational setups would be very useful. A reason is that while the Ryu-Takayanagi (RT) \cite{Ryu:2006ef} surface is an imaginary surface, the holographic dual of the replicated geometry drastically changes the structure of spacetime. Intuitively, we expect the dual geometries (if they exist) to have a smooth horizon where the RT surface was and to be similar to the original geometry far from the RT surface. We expect that universal results like the ones we have proven could aid in understanding what the dual geometries could be or how to compute R\'enyi entropy holographically in general. It may be particularly tractable to study this question near $n=1$; exploring the relation \eqr{KK} seems like a plausible and worthwhile target.

\section*{Acknowledgments}

We thank Joan Camps, Horacio Casini, Sean Hartnoll, Juan Maldacena, Gim-Seng Ng, Mukund Rangamani, Vladimir Rosenhaus, Ben Safdi and Alexander Zhiboedov for helpful discussions.
 We also thank Mark Mezei for comments on a draft. E.P. wishes to thank the Michigan Center for Theoretical Physics and the Aspen Center for Physics for hospitality during this work, which was supported in part by National Science Foundation Grant No. PHYS-1066293. E.P. has also received funding from the European Research Council under the European Union's Seventh Framework Programme (FP7/2007-2013), ERC Grant agreement STG 279943, “Strongly Coupled Systems”.  A.L. acknowledges support from ``Fundacion La Caixa".

\appendix 
\section{Details of first order $d=6$ integral}

Our starting point is \eqr{6dpert}. We consider only the transverse components, $h^{\t}_{\t}-{1\over 5}h^r_r$. The only terms that give a log from the transverse integral are of $O(r^4)\sim O(x^4)$; extending the adapted metric to this order and keeping only the term linear in curvature, one finds (see for example equation (2.16) of \cite{AlvarezGaume:1981hn})
\begin{equation}
h_{ac} =-{1\over 20} \p_e\p_f R_{abcd}x^ex^fx^bx^d~.
\end{equation}
As we are only looking at components diagonal in $(ac)$, it is clear that  $x^bx^d$ collapses to $\delta^{bd}r^2$ times some angular pieces. This then forces $x^ex^f$ to do the same, which leads to the Laplacian. Proceeding straightforwardly, one arrives at
\begin{eqnarray*}
\int^{2\pi}_0 d\t (h^{\t}_{\t}-{1\over 5}h^r_r ) =  -{\pi\over 40}r^4\delta^{ac}\delta^{bd} (\p_1^2 + \p_2^2) R_{abcd}~.
\end{eqnarray*}
Plugging into (the transverse part of) \eqr{6dpert}, and equating with the known CFT expression \eqr{6dpertcft}, we get
\begin{equation}
8\pi f_{B_3}(n) \int \nabla^2 C^{ab}_{~~ab} = -{\pi\over 40}{n\over 2(1-n)} F(n) \int (\nabla^2 R^{ab}_{~~ab}+...)~.
\end{equation}
Under the integral, $\int_{\S} \nabla^2_{y_i} (...)=0$. From the definition of the Weyl tensor, one can show that
\begin{equation}\la{weylriem}
C^{a b}_{~~ a b}=\frac{d-3}{d-1} R^{a b}_{~~ a b}+...
\end{equation}
Using \eqr{weylriem} in $d=6$ leads to the relations \eqr{fb3F} and \eqr{fb3fa}. (We have only computed the transverse piece, and are assuming that the parallel components of the integral yield the remaining terms on the right-hand side of \eqr{weylriem}.)

To check the overall coefficient in \eqr{fb3fa}, we evaluate both sides at $n=1$ and check that $f_{B_3}(1)=B_3$. From  \eqr{fs}, we have
\begin{eqnarray}
F(1) &=& -{(n-1)\over \pi \Vol(\IH^{d-1})}S'_{n=1}(\S=S^{d-2})~.
\end{eqnarray}
It was shown in \cite{Perlmutter:2013gua} that
\be
S'_{n=1}(\S=S^{d-2}) = -{\rm Vol}(\IH^{d-1})\cdot {\pi^{d/2+1}\Gamma(d/2)(d-1)\over(d+1)!}C_T ~.
\ee
$C_T$ is defined via the stress tensor vacuum two-point function in flat space  \cite{Osborn:1993cr} 
\be
\lang T_{ab}(x)T_{cd}(0)\rang_{\IR^d} = C_T{I_{ab,cd}(x)\over x^{2d}}~.
\ee
$I_{ab,cd}(x)$ is a particular tensor structure which can be found in \cite{Osborn:1993cr}. Thus, in $d=6$,
\begin{eqnarray}
F(1) &=&(n-1)\pi^3 {10\over 7!}C_T
\end{eqnarray}
and from \eqr{fb3F} and \eqr{fb3fa}, we have
\begin{equation}
{f_{B_3}(1)} = {\pi^3\over 64\cdot 7!} C_T\cdot {5\over 3}~.
\end{equation}
The ratio $B_3/C_T$ is a theory-independent quantity. Using results in the literature for the free (2,0) tensor multiplet,\footnote{See e.g. equations (2.29) of \cite{Bastianelli:2000hi} and (3.23) of \cite{Bastianelli:1999ab}, where they use the same normalization that we use for the Weyl anomaly.} one sees that indeed, $f_{B_3}(1)=B_3$ as desired.

\section{Some additional comments about the R\'enyi perturbative expansion}

\subsection{Effect of conformal transformations}

Conformal perturbation theory around a given background will not be conformally invariant unless we allow for coordinate-dependent couplings. 

If we consider two conformally related metrics $ds_{\Omega}^2=\Omega^{-2} ds^2$, then when we deform the original theory by an operator (we consider a conformal primary for simplicity)
\begin{equation}
\delta S=\int \sqrt{g} \lambda O=\int_{{\cal M}} \sqrt{g_{\Omega}} \Omega^d \lambda \Omega^{-\Delta} O_{\Omega}=\int_{{\cal M}} \sqrt{g_{\Omega}} \lambda_{\Omega}  O_{\Omega} ~~ ,
\end{equation}
the perturbative expansions are the same if we identify the coordinate dependent coupling $\lambda_{\Omega}=\Omega^{d-\Delta} \lambda$. If the deformation is $\frac{1}{2} T_{\mu \nu} h^{\mu \nu}$,  this means that we keep $h$ fixed.\footnote{Of course if we change the metric the theory is still conformal, so all terms in conformal perturbation theory will be conformally invariant.} Alternatively, if we want to rescale $h$, the coupling becomes $h^{\mu \nu}=h^{\mu \nu}_{\Omega} \Omega^{-2}$.

\subsection{Recovering the entanglement entropy}

Now we would like to compute the EE from  expression \eqr{sum} for the R\'enyi entropy.  For simplicity of notation, it is easier to formally work at finite coupling and do the perturbative expansion later. That is we can rewrite equation \eqr{sum} in the following way
\begin{equation}
\partial_g S_n=\frac{n}{1-n} \int_{\M_1} ( \langle O \rangle_{g,n}-\langle O \rangle_{g,n=1}) \quad \Rightarrow\quad 
\partial_g S_{EE}=-\partial_n  \int_{\M_1} \langle O \rangle_{g,n}
\end{equation}

It is trivial to see that this expression reproduces the first order correction.

If we want to go to second order, then we take another $g$-derivative and set $g=0$. There are two contributions to the $n$-derivative: one contributing from the expectation value $\langle O O \rangle_{n}$ and the other one from the integral $\int_{{\cal M}_n}$.

First, we can easily see what the contribution by taking first the $n$-derivative. We can do this, because at finite $g$ the perturbation is shifting the modular Hamiltonian by $K_g-K_0=-g \int_{\M_1} O$, see also \cite{RS3}. (Note that away from $g=0$, different conformal frames won't be equivalent and while $T_{\tau \tau}$ will still be performing the replica trick it is not clear that it is also computing the entanglement entropy in the respective frame.) Then the $n$-derivative just lowers the finite $g$ modular Hamiltonian:
\es{}{
-\partial_g \partial_n \langle O \rangle_{g,n}=\partial_g \langle O K_g \rangle|_{g=0}&= \langle O \partial_g K_g \rangle |_{g=0}+\langle O K_0 \int_{\M_1} O \rangle\\&=-\langle O \int_{\M_1} O \rangle+\langle   K_0O \int_{\M_1} O \rangle~.
}
  
  We can also do the derivatives in the opposite order. A particularly easy way to do the $n$-derivative of the expectation value is by using coordinates  $\tau'$ such that the $g_{\tau' \tau'}=n^2$, and then taking the derivative inside the integral (without deriving the operators with respect to $n$).  One can see then that $-\partial_n \langle O(0) O(\tau') \rangle_n=\langle K O(0) O(\tau')  \rangle_n$. The rest follows 
\es{}{
  -\partial_n \partial_g \langle O \rangle_{g,n}&=-\partial_n  \langle O(0)   \int_0^{2\pi} n d\tau' \int O(\tau') \rangle_n|_{n=1}\\&=  -\langle O \int_{{\cal M}}O \rangle+ \langle K_0 O \int_{\cal M} O \rangle
  }
  which is the same result obtained above.\footnote{Note that at $n=1$, one also gets the same answer working in the usual $\tau$ coordinates and taking the derivative of the integral \emph{after} setting $n=1$ in the correlator:	  $-\partial_n [\langle O(0) \int_{{\cal M}_n} O(\tau') \rangle_{n}]|_{n=1}+\partial_n \langle O(0) \int_{{\cal M}} O(\tau') \rangle_{n}|_{n=1}=-\partial_n \langle O(0) \int_{{\cal M}_n} O(\tau') \rangle_{n=1}|_{n=1}=-\langle O(0) \int_{{\cal M}} O(\tau') \rangle_{n=1}$. This seems the proper way of analytically continuing the integral in $n$.} In the previous case one could wonder if the reasoning involving $K_g$ was also correct for a geometric perturbation: if we think of Rindler, then if we perturb the metric it is not clear that one can still talk about a deformed Rindler Hamiltonian. However, from this point of view we see that there should be no problem. 
  
\section{The relation between $f_b(n)$ and $f_c(n)$: Details}
The purpose of this appendix is to try to unravel the structure of the sum \eqr{int4}. In doing so, we will argue that the relation between $f_b(n)$ and $f_c(n)$ boils down to understanding the dynamics of marginal operators $O_4$ in the CFT.

\vs
Let us focus on the stress tensor contribution to \eqr{int4}. This will take the same functional form for all CFTs. In \eqr{FnR}, we wrote down the expectation value of the $(yy)$ components of the stress tensor on $\C_n\times \IR^2$. In the language of \eqr{onept},
\es{}{a_T(n) = -{1\over 3}F(n)~.}
Plugging this into \eqr{int4}, we see that the $n$-dependence simply equals $f_c(n)$, as defined in \eqr{fcF}. So the stress tensor contributes to \eqr{int4} as
\es{c2}{\left({f_c(n)\over 6\pi}\int d^2y K^a_{ij}K^a_{kl} {\cal T}^{ijkl}_{T}\right)\log R/\eps~.
}
This is satisfyingly similar to the contribution $\d^{(1)}S_n|_{K^2}$ we derived in \eqr{metric2KK}. 

We now wish to argue that $f_b(n)=f_c(n)$ in theories for which this is the {\it only} contribution to the sum \eqr{int4}. To do so, let us take the $n\rar 1$ limit: that is, we consider the second order perturbation of the EE. In this limit, the computation above requires us to compute $\p_n\langle T^{ij}(z') T^{kl}(z)\rangle|_{n=1} =-  \langle T^{ij}(z') T^{kl}(z) K_{\rm Sph}\rangle_1$. Performing the OPE between the two stress tensors, the only term that survives the expectation value is the stress tensor itself, because $\lang O K_{\rm Sph} \rang_1=0$ for $O\neq T$.\footnote{An exception to this rule is for $K_{\rm Sph}$ defined in the free conformal scalar theory. In this case, $K_{\rm Sph}$ is built from a non-conformal stress tensor that includes a total derivative term inherited from the boundary of the singular cone \cite{Lee:2014x}.} 

Now, when $n\rar 1$ we must obtain the known result,
\es{}{
S_{EE}\big|_{K^2} = -\left({c\over 2\pi}\int_{\S}\big( {K^a_{i j}} {K^a_{j i}} -\frac{1}{2} (K^a_{ i i})^2\big)\right)\log R/\eps~.
}
This is simply \eqr{fursaev} with $f_b(1)=f_c(1)=c$. If this is to follow from \eqr{c2} at $n=1$, the tensor ${\cal T}^{ijkl}_{T}$ must have a particular index structure and overall normalization. As non-trivial as this is to prove, the crucial point is that this tensor is independent of $n$. Therefore, the generalization to finite $n$ would amount simply to the replacement $c \rar f_c(n)$: that is to say, $f_b(n)=f_c(n)$, up to contributions to $\d^{(2)}S_n|_{K^2}$ from $\D=4$ operators besides the stress tensor. 

Before moving on, let us give a realistic look at the tensor ${\cal T}^{ijkl}_{T}$. Hiding therein is the OPE coefficient $c_{TTT}$, which is a sum of three independent, conformally invariant tensor structures, each multiplying a constant. Only one linear combination of these is proportional to $c$. This seems to imply that upon integration, only the structure proportional to $c$ survives. The resulting term must also have the correct index structure such that, when contracted with the extrinsic curvatures, it yields the conformally invariant structure in \eqr{51}. As inspiration, we wish to highlight an apparently similar result in Section 7 of  \cite{Osborn:1993cr} that may be useful.

We now consider other operators $O_4\neq T$. Using arguments in Section 5.1 and the unitarity bound $\D\geq s+2$ for operators with spin $s$, the sum over operators $O_4\neq T$ localizes further onto scalar operators, and onto symmetric, traceless spin-2 currents that are not proportional to the stress tensor. While this class of operators does not contribute at $n=1$, one cannot rule out its contribution at finite $n$. 

However, in a wide class of theories, such contributions do not exist. That is, in the following theories, there exist neither $\D=4$ scalar operators, nor other symmetric, traceless spin-2 currents, that appear in the $TT$ OPE and have nonvanishing expectation values on the conical spacetime.

\vs
\bul {\bf Free theories}

\vs

  For the free scalar, we have $O_4=\lbrace \phi^4, \L\rbrace$, where $\L$ is the Lagrangian operator. $\phi^4$ has vanishing three-point function with the stress tensor\footnote{$\langle T T \phi^4 \rangle=\langle T \phi^2 \rangle^2=0$} and $\L=(\partial \phi)^2$ doesn't contribute because it is a descendant of $\phi^2$, which has a constant expectation value $\lang \phi^2\rang_n$.

For the free fermion, we have $O_4=\L=\bar{\psi}\gamma\cdot \partial \psi$. The expectation value of this operator will be proportional to $\frac{\gamma x_a(\tau)}{r^5}$, so it vanishes by parity.

For the free Maxwell field, the only gauge-invariant operator is $O_4=\L$. As shown in  \cite{Osborn:1993cr}, $c_{TT\L}=0$. 

In the scalar and fermion theories, $f_b(n)=f_c(n)$ was shown directly in \cite{Lee:2014xwa}. Our arguments are consistent with that, and extend the equality to the case of a free Maxwell field. 

\vs
\bul {\bf Planar $\N=4$ SYM and other holographic CFTs}

\vs

Consider planar $\N=4$ SYM. The only $\D=4$ operator at finite $\l$ which is not charged under the $SO(6)_R$ symmetry 
is the Lagrangian density. It is known that $c_{TT\L}=0$ for all $\lambda$. One way to see this is to resort to the $U(1)_Y$ ``bonus'' symmetry of the strongly coupled theory, which forces all non-singlet correlators to vanish \cite{Intriligator:1998ig,Eden:1999gh}. $T$ and $\L$ have $U(1)_Y$ charges zero and four, respectively.\footnote{We thank Alexander Zhiboedov for pointing this out.} From this it follows that $c_{TT\L}=0$ at {\it all} $\l$, because all three-point functions of protected operators are unrenormalized. 

Gravitationally, $c_{TT\L}=0$ maps to the statement that the dilaton does not have a nonzero three-point vertex with two gravitons. Furthermore, the statement that an expectation value for other scalar operators would break the R-symmetry maps to the fact that all other bulk scalars besides the dilaton (descending from KK modes) can be set to zero in the hyperbolic black hole background. 

These last two arguments can be generalized to other strongly coupled CFTs besides $\N=4$ SYM. Consider a CFT that admits a holographic limit. This CFT will satisfy the previous conditions if the bulk theory possesses either of the following two properties: a) no cubic vertex between two gravitons and a massless scalar, or b) a continuous family of hyperbolic black hole solutions that do not turn on a massless scalar. 

In particular, replacing the $S^5$ in the type IIB supergravity compactification to $D=5$ with another transverse manifold will not affect the previous conclusions. 

\vs
\bul {\bf$\N=4$ SYM}
\vs

Finally, let us return to $\N=4$ SYM, but now for arbitrary gauge group and complexified gauge coupling $\t = {\theta_{YM}\over 2\pi} + 4\pi i g_{YM}^{-2}$. Even in this case, the previous conclusions hold. Let us give two arguments. In what follows, we ignore unprotected operators, which acquire anomalous dimensions and hence cannot be exactly marginal.

First, we again note that of the protected operators, the only $SO(6)_R$ singlet scalar operator is the Lagrangian. Using our earlier arguments for free fields, $c_{TT\L}=0$ at the free fixed point. This three-point function is protected against renormalization as a function of $\t$. One way to see this is that the $\t$-derivative of the stress tensor two-point function $\langle TT \ra$ must vanish. This derivative yields a spacetime integral over the three-point function $\langle TT\L \ra$. The integral over the lone tensor structure (see e.g. \cite{Osborn:1993cr}) is non-vanishing, so the OPE coefficient must vanish: that is, $c_{TT\L}=0$ for all $\t$. 

An alternative argument again utilizes the bonus symmetry of $\N=4$ SYM, in particular the conjecture that OPE coefficients obey $U(1)_Y$ selection rules even away from strong coupling and the planar limit \cite{Intriligator:1998ig,Intriligator:1999ff}. It was argued in \cite{Intriligator:1999ff} that for arbitrary gauge group and $\t$, the only non-vanishing OPE coefficients for which at least two of the three operators sit in protected multiplets of the $\N=4$ superconformal algebra are $U(1)_Y$ singlets. This again implies that only $U(1)_Y$ singlets can appear in the $TT$ OPE; but as we already noted, this rules out the Lagrangian.

Finally, we note that our first argument above rules out the contribution of the Lagrangian to our method of defining $f_b(n)$ in all CFTs which possess a marginal gauge coupling. If, as in $\N=4$ SYM, the Lagrangian is the only exactly marginal scalar operator that can appear in the $TT$ OPE, then $f_b(n) = f_c(n)$ in such theories as well.

\bibliographystyle{utphys}
\bibliography{biblio}

\end{document}